\documentclass[]{article} 

\usepackage[utf8]{inputenc} 
\usepackage[T1]{fontenc}
\usepackage[nonatbib, final]{neurips_2019}
\usepackage[backend=bibtex]{biblatex}
\usepackage{tikz}
\usetikzlibrary{shapes,arrows,chains}
\usetikzlibrary[calc]

\usepackage{mathtools, amsmath, amssymb, amsfonts, amsthm, mathtools, mathabx}
\usepackage[]{authblk}
\usepackage{hyperref}

\usepackage{booktabs}
\usepackage{multirow}

\usepackage{algorithm}
\usepackage{algpseudocode} 

\usepackage{xcolor}
\usepackage{wrapfig}

\usepackage[font=small,skip=5pt]{caption}

\newcommand{\R}{\mathbb{R}}
\newcommand{\U}{{U}}
\newcommand{\X}{{X}}
\newcommand{\Y}{{Y}}
\newcommand{\Z}{{Z}}
\newcommand{\px}{q}
\newcommand{\pz}{p}

\DeclareMathOperator*{\KL}{KL}

\usepackage{tikz}
\usetikzlibrary{backgrounds}
\usetikzlibrary{arrows}
\usetikzlibrary{shapes,shapes.geometric,shapes.misc}

\tikzstyle{tikzfig}=[baseline=-0.25em,scale=0.5]

\pgfkeys{/tikz/tikzit fill/.initial=0}
\pgfkeys{/tikz/tikzit draw/.initial=0}
\pgfkeys{/tikz/tikzit shape/.initial=0}
\pgfkeys{/tikz/tikzit category/.initial=0}

\pgfdeclarelayer{edgelayer}
\pgfdeclarelayer{nodelayer}
\pgfsetlayers{background,edgelayer,nodelayer,main}

\tikzstyle{none}=[inner sep=0mm]

\newcommand{\tikzfig}[1]{%
{\tikzstyle{every picture}=[tikzfig]
\IfFileExists{#1.tikz}
  {\input{#1.tikz}}
  {%
    \IfFileExists{./figures/#1.tikz}
      {\input{./figures/#1.tikz}}
      {\tikz[baseline=-0.5em]{\node[draw=red,font=\color{red},fill=red!10!white] {\textit{#1}};}}%
  }}%
}

\newcommand{\ctikzfig}[1]{%
\begin{center}\rm
  \input{#1.tikz}
\end{center}}

\tikzstyle{every loop}=[]

\tikzstyle{dashezz}=[-, dashed]
\tikzstyle{dlvs}=[-, fill={rgb,255: red,255; green,215; blue,215}, draw=black]
\tikzstyle{calls}=[-, fill={rgb,255: red,255; green,235; blue,235}]
\tikzstyle{predlvs}=[-, fill={rgb,255: red,255; green,195; blue,195}]
\tikzstyle{encoder}=[-, fill={rgb,255: red,135; green,135; blue,255}]
\tikzstyle{compression}=[-, fill=none]
\tikzstyle{return}=[-, fill=none]
\tikzstyle{arrow}=[->]
\tikzstyle{noise}=[-, fill=none]
\tikzstyle{yt}=[-, fill=none]

\AtBeginBibliography{\footnotesize}

\makeatletter
\renewcommand*{\eqref}[1]{%
  \hyperref[{#1}]{\textup{\tagform@{\ref*{#1}}}}%
}
\makeatother

\title{Multi-Asset Spot and Option Market Simulation}
\author[1, 2]{Magnus Wiese\protect\thanks{Corresponding author: \texttt{magnus.wiese@jpmorgan.com}}\ \ }
\author[1]{Ben Wood}
\author[1]{Alexandre Pachoud}
\author[2]{\\Ralf Korn}
\author[1]{Hans Buehler}
\author[1]{Phillip Murray}
\author[1]{Lianjun Bai}
\affil[1]{J.P. Morgan\thanks{Opinions expressed in this paper are those of the authors, and do not necessarily reflect the view of J.P. Morgan.}\protect\\{\small 25 Bank Street, London, United Kingdom}\protect\\ ~}
\affil[2]{University of Kaiserlautern\protect\\ {\small Gottlieb-Daimler Stra{\ss}e 48, Kaiserslautern, Germany}}

\usepackage{mathtools}
\usepackage{amsthm}

\newcommand{\N}{\mathbb{N}}
\renewcommand{\R}{\mathbb{R}}

\newcommand{\bfu}{\mathbf{u}}

\newcommand{\bfx}{\mathbf{x}}
\newcommand{\bfz}{\mathbf{z}}
\newcommand{\bfy}{\mathbf{y}}

\newcommand{\baru}{\bar{\bfu}}
\newcommand{\barx}{\bar{\bfx}}
\newcommand{\bary}{\bar{\bfy}}
\newcommand{\barz}{\bar{\bfz}}

\newtheorem{thm}{Theorem}

\theoremstyle{definition}
\newtheorem{remark}{Remark}

\begin{document}
\maketitle

\begin{abstract}
	We construct realistic spot and equity option market simulators for a single underlying on the basis of normalizing flows. We address the high-dimensionality of market observed call prices through an arbitrage-free autoencoder that approximates efficient low-dimensional representations of the prices while maintaining no static arbitrage in the reconstructed surface. Given a multi-asset universe, we leverage the conditional invertibility property of normalizing flows and introduce a scalable method to calibrate the joint distribution of a set of independent simulators while preserving the dynamics of each simulator. Empirical results highlight the goodness of the calibrated simulators and their fidelity.
\end{abstract}

\section{Introduction}
There is a great interest in applying deep reinforcement learning \cite{sutton2018reinforcement} to financial markets \cite{buehler2019deep,buehler2019deep_rl_version,deng2016deep, nevmyvaka2006reinforcement}. Unfortunately, on a daily time scale, the amount of market data is not sufficient to train reinforcement learning algorithms for single or multiple underlyings: ten years worth of daily stock data amount to a few thousand samples. Thus, making it impossible to train reinforcement learning algorithms to generalise to market dynamics without overfitting. Due to this data scarcity challenge, there is increasing attention in finding new ways of modeling the dynamics of financial markets by using (deep) generative modeling \cite{assefa2020generating, Horvath2020, cohen2021arbitrage, cuchiero2020generative, pardo2020mitigating, de2021tackling, kondratyev2019market, koshiyama2019, Marti_2020, Hao2020, ni2021sigwasserstein, van2021monte, ruf2020neural, Wiese2019, QuantGANs}.

In this paper, we progress our work on modeling single-asset equity option market simulators with neural networks \cite{Wiese2019} by introducing a novel flow-based market simulator. Moreover, we extend them to the multi-asset case for more realistic downstream applications such as \citetitle{buehler2019deep} \cite{buehler2019deep} for multi-asset portfolios and / or basket options. Just as in the single asset world a new market simulator needs to be calibrated that models the joint dynamics of the portfolio's underlyings. Unfortunately, after performing a quick combinatorial exercise, it becomes evident that for an $N$-variate asset universe the total number of portfolio combinations that can be constructed is $2^N$. This makes it practically not possible to use multi-asset market simulators when the asset universe is large since a new market simulator would have to be calibrated for every new combination. Although one could be tempted to create a single model that generates the whole asset universe in practice this is not feasible due to liquidity constraints, i.e. not-available underlyings limiting the size of the dataset to a few or no dates.

\subsection{Main results}
As a result of this scalability issue we propose a novel \emph{conditionally invertible} single asset market simulator $ T_\theta: \X^p \to (\Z \stackrel{\cong}{\to} \X) $ based on normalizing flows \cite{papamakarios2019normalizing}. The simulator map $T_\theta$ relates random adapted i.i.d. noise $\bfz_{t+1} \sim p$ (the innovation) and the past $p$-lagged market states $\bfy_t = (\bfx_{t}, \dots, \bfx_{t-p+1})$ to the next day's market state $\bfx_{t+1}$
\begin{equation*}
	\bfx_{t+1} = T_\theta(\bfz_{t+1}; \bfy_t) \sim \px_\theta
\end{equation*}
and allows for fixed $\bfy_t, \bfx_{t+1} \in \X $ for efficient computation of the \emph{latent} time series $\bfz_{t+1} = T^{-1}_\theta(\bfx_{t+1}; \bfy_t)$. In our construction, the market state $\bfx_t$ represents the spot's log-return and an arbitrage-free low-dimensional efficient representation of the call price surface through the use of \emph{autoencoders} \cite{hinton2006reducing} and \emph{discrete local volatilities} \cite{buehler2017discrete} (DLVs). 

For a set of $N$ simulators $T^i_\theta: \X^p \to (\Z \stackrel{\cong}{\to} \X), i=1, \dots, N$ calibrated on $N$ distinct underlyings, we leverage the invertibility property of the flow to compute the \emph{latent} time series of each underlying
\begin{equation*}
	\bfz_{t+1}^i = (T^i_\theta)^{-1}(\bfx^i_{t+1}; \bfy^i_{t}), \quad i = 1, \dots, N
\end{equation*}
which allows us to study and model the joint density of the \emph{joint} latent process $\barz_t = (\bfz^1_t, \dots, \bfz^N_t)$ by employing well-studied copulas \cite{nelsen2007introduction}. Using copulas for modeling the joint is advantageous: there is a clear trade-off between complexity and scalability, it does not change the marginal density of the latent process $\bfz^1, \dots, \bfz^N \sim \pz$, and hence, does not change the dynamics of the individual simulators. Our numerical results highlight that utilizing the Gaussian copula lead to a realistic generation of the multi-asset market's returns.    

\subsection{Contributions} Our contributions are as follows:
\begin{itemize}
	\item We formulate the single asset market simulator such that simulating spot and the equity option market without static arbitrage boils down to a pure dimensionality-reduction and generative modeling problem under the real-world measure $\mathbb{P}$. Removal of statistical arbitrage and calibration of the change-in-measure from $\mathbb{P}$ to the minimal equivalent near-martingale measure $\mathbb{Q}$ is addressed in \cite{buehler2021deep}. 
	\item Due to the high-dimensionality of the market observed prices, we introduce a DLV-based autoencoder which is able to learn an efficient low-dimensional representation of the call price surface while simulataneously guaranteeing no static arbitrage in the reconstructed call price surface. 
	\item We show that single asset market simulators can be efficiently approximated by using the Kullback-Leibler divergence. Thereby, we reduce costly black-box adversarial calibration \cite{arjovsky2017wasserstein, goodfellow2014generative, gulrajani2017improved} to fast supervised learning with tractable conditional densities. 
	\item We show numerically that well-studied copulas can model the joint density of multiple distinct market simulators for Eurostoxx 50 and S\&P 500. 
\end{itemize}

\subsection{Outline} Our paper is structured as follows. We introduce spot and equity option markets in \autoref{sec:setup}. A review of related work and literature on generative modeling of financial markets is provided in \autoref{sec:literature_overview}, while \autoref{sec:normalizing_flows_background} contains the introduction of normalizing flows. Sections \ref{sec:market_simulator} and \ref{sec:multi_simulator} present formally the spot and equity option market simulator for a single underlying and multiple underlyings. Both sections \ref{sec:market_simulator} and \ref{sec:multi_simulator} provide numerical results that demonstrate the performance of the algorithm. The paper is concluded in \autoref{sec:conclusion} and highlights directions of future research. 

\begin{figure}[!h]
	\centering
	\includegraphics[width=\textwidth]{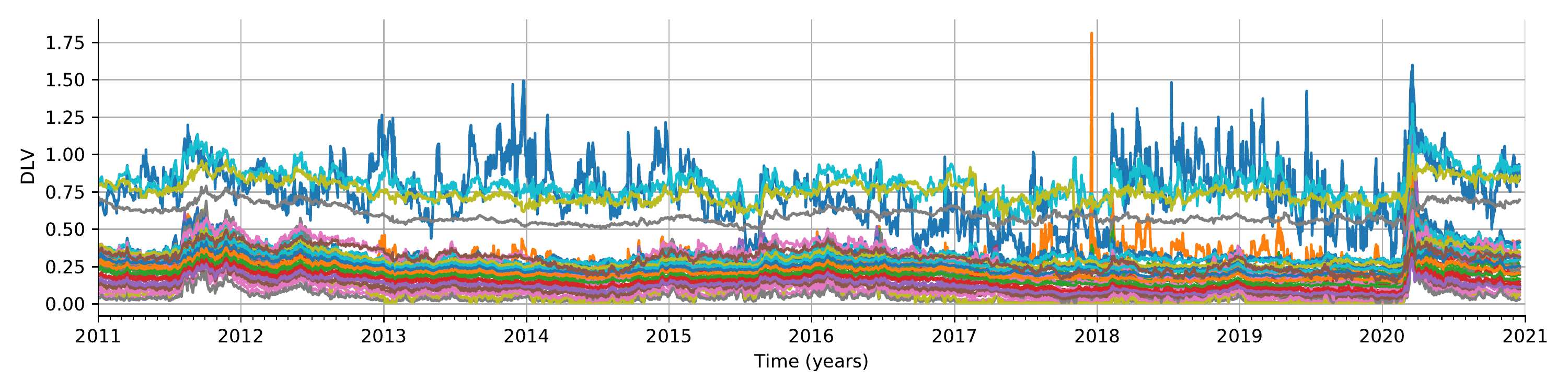}
	\caption{DLVs of the Eurostoxx 50 for Jan 2011 - Nov 2021 for the grid maturities and relative strikes respectively $\lbrace 20, 40, 60, 120 \rbrace \times \lbrace 0.80, 0.85, 0.90, \dots, 1.20 \rbrace$.}
	\label{fig:dlvseurostoxx}
\end{figure}

\section{Spot and equity option markets}
\label{sec:setup}
Our goal is to simulate the dynamics of the equity spot and call / put option market. The spot price is the current price in a marketplace at which an asset can be bought or sold. Here we restrict ourselves to indices and stocks. A call (put) option is a financial instrument that gives the holder the right, but not the obligation, to buy (sell) the underlying asset for a predetermined price on a predetermined date. The payoff of a vanilla option for relative strike $k \geq 0$ and maturity $t \geq  0$ can be written as $\max(0, \alpha(S_t-k))$ where $S_t$ is the spot price at time $t$, and $\alpha = 1$ defines the call payoff and $\alpha=-1$ defines the put payoff. 

\subsection{Guaranteeing no static arbitrage with discrete local volatilities}
\label{sec:no_static_arb}

\begin{wrapfigure}[14]{r}{0.5\textwidth}
	\centering
	\includegraphics[width=\linewidth]{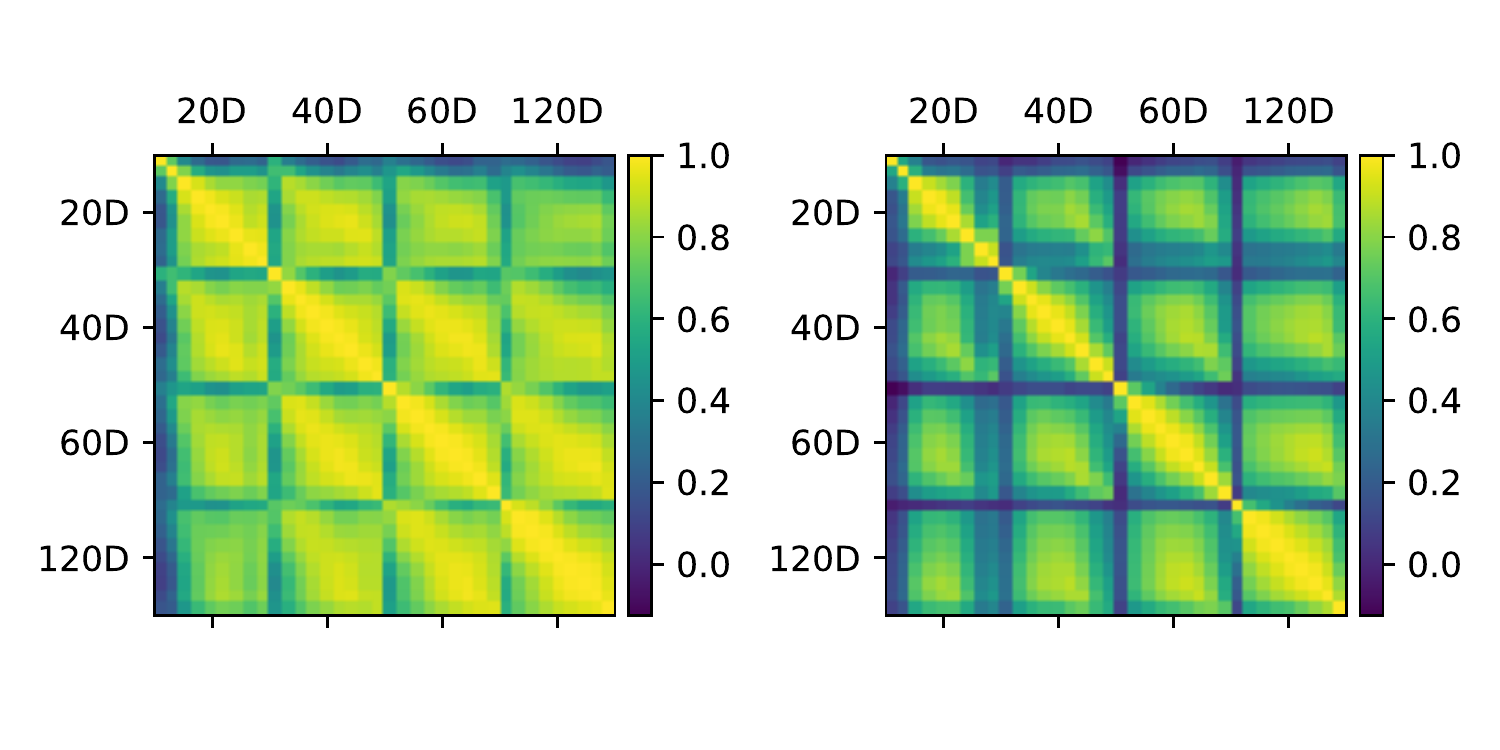}
	\caption{Cross correlation matrix of Eurostoxx 50 DLV levels (left) and their returns (right).}
	\label{fig:dlvs_correlated}
\end{wrapfigure}

On the option market vanilla call and put options cannot only be exchanged for a single maturity-strike pair, but for a whole list of maturity and strike combinations. Listed options need to satisfy ordering constraints (see \cite{buehler2017discrete, cohen2021arbitrage, schweizer2008arbitrage}) otherwise  prices will offer the opportunity to make a risk-free profit, more commonly referred to as \emph{static arbitrage}. Although static arbitrage can be observed in financial markets its presence is spurious, difficult to model and easily exploitable by a downstream algorithm such as Deep Hedging. We therefore, construct a market simulator that generates call prices that satisfies no static arbitrage by employing \emph{discrete local volatilities} (\emph{DLVs}) \cite{buehler2017discrete, buehler2018discrete} as an arbitrage-free representation of a grid of call prices. 

For completeness, we briefly review the construction of DLVs and will assume that we are given a grid of market call prices for simplicity. Interpolation of a lattice is discussed in \cite{buehler2017discrete,buehler2018discrete}. Consider $N$ market observed call options $\mathbb{C}^{j, i}$ with maturities $0<\tau_1<\dots<\tau_m$ and relative strikes $0<x_{-1}<\dots<1<\dots<x_{n}$ defined around the unit forward, i.e. we look at strikes that are normed by $F(\tau_{j})$, the forward price of the underlying stock. We assume that call options with relative strikes at the boundary are priced at their intrinsic value, i.e. $C^{j, -1} = 1 - x_{-1}$ and $C^{j, n} = 0$. The optimisation problem is to find the closest surface $C$ to the market observed lattice $\mathbb{C}$ satisfying the linear arbitrage-free constraints (see \cite{buehler2017discrete,buehler2018discrete}): 
\begin{enumerate}
	\item $C^{j, 0} \geq 1 - x_{j}$,
	\item $C^{j, n_j - 1} \geq 0$,
	\item $ \Gamma^{j, i} \coloneqq \frac{\Delta^{j, i}  - \Delta^{j, i-1}}{\frac{1}{2}(x_{i+1} - x_{i})} \geq 0 $ where (upward) delta is defined as $\Delta^i_j = \frac{C^{j, i+1} - C^{j, i}}{\frac{1}{2}(x_{i+1} - x_{i})}$,
	\item $ b\Theta^{j, i} \coloneqq C^{j, i} - C^{j-1, i} \geq 0 $, and impose the constraint 
	\begin{equation*}
		\sigma_{-}^2\frac{1}{2}\Gamma^{j, i} x_j^2 (\tau_j - \tau_{j-1})\leq b\Theta^{j, i} \leq \sigma_+^2 \frac{1}{2}\Gamma^{j, i} x_j^2 (\tau_j - \tau_{j-1})
	\end{equation*}
\end{enumerate}
These constraints are feasible by construction. The DLV, defined as 
\begin{equation*}
	\Sigma^{j, i} \coloneqq \sqrt{2\dfrac{b\Theta^{j, i}}{\Gamma^{j, i} x_j^2 (\tau_j - \tau_{j-1})}},
\end{equation*}
is therefore well-defined and bounded from below and above by $\sigma_-$ and $\sigma_+$. Note that the mapping from the arbitrage-free surface $C$ to the DLVs $\mathbf{\Sigma} = (\Sigma^{j, i})_{i, j}$ is bijective. For illustration we display the DLVs of Eurostoxx 50 for the January 2011 to November 2021 in \autoref{fig:dlvseurostoxx} for the grid of maturities $\mathcal{T} = \lbrace 20, 40, 60, 120 \rbrace$ and relative strikes  $\mathcal{K}= \lbrace 0.80, 0.85, 0.90, \dots, 1.20 \rbrace$. 

\subsection{Stylized facts of spot and equity option markets}
\label{sec:stylized_facts}
To assess the fidelity of a calibrated market simulator, we review the features of the spot and equity option market. Spot market dynamics have been extensively studied in the past. The most characteristic features of the spot market were summarised by \citeauthor{cont2001empirical} \cite{cont2001empirical} and coined as \emph{stylized facts}. We summarise the most prominent features of spot log-returns from \citetitle{cont2001empirical} \cite{cont2001empirical} as follows: 
\begin{itemize}
	\item \textbf{Fat tails}: evaluation of the tails of the empirical unconditional return distribution indicates the presence of \emph{fat tails}, i.e. power-law decay in the tail of the density. 
	\item \textbf{No serial correlation}: spot log-returns have close to zero serial autocorrelation, i.e. they are hard to predict linearly. 
	\item \textbf{Volatility clusters}: volatility tends to appear in clusters. Large absolute spot log-returns generally follow large absolute spot log-returns. This observation can be measured by computing the autocorrelations of absolute or squared spot log-returns. 
	\item \textbf{Leverage effects}: absolute spot log-returns and log-returns are negatively correlated: a large fall in spot levels causes an increase in volatility.  
\end{itemize}
The empirical observations above influenced the design of stochastic volatility models in the past by incorporating these observed dynamics, such as leverage effects and volatility clustering, into the model (for example see the Heston model \cite{heston1993closed}). 

\begin{figure}[!htb]
    \centering
    \begin{minipage}{.5\textwidth}
        \centering
        \includegraphics[width=\textwidth]{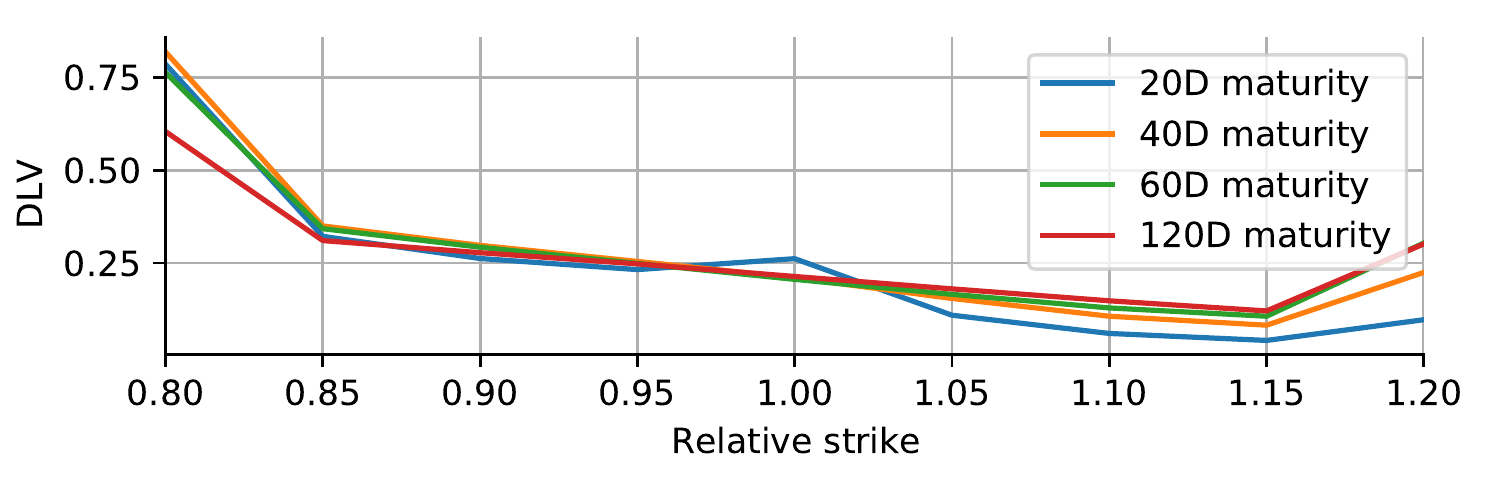}
		\caption{Eurostoxx 50 volatility smile.}
        \label{fig:dlv_smile}
    \end{minipage}%
    \begin{minipage}{0.5\textwidth}
        \centering
		\includegraphics[width=\textwidth]{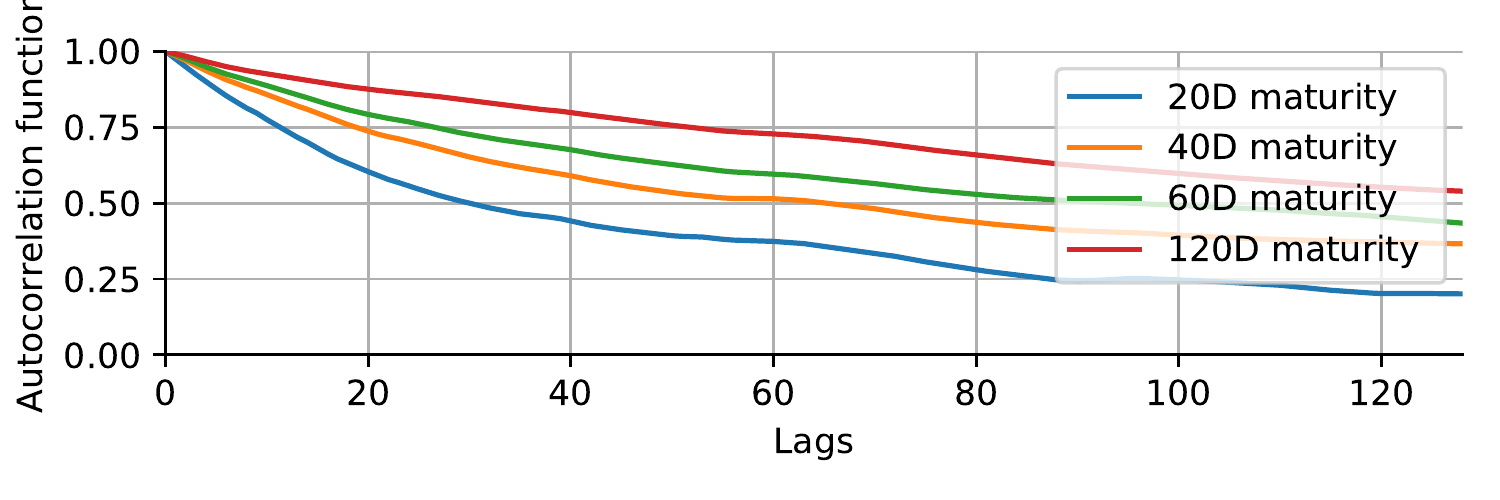}
		\caption{Autocorrelation of ATM DLVs.}
        \label{fig:dlvs_atm_acf}
    \end{minipage}
\end{figure}

Stylized facts of DLVs are discussed for the first time in this paper, but due to their local volatility inspired parametrisation are consistent with general observations of equity option volatility surfaces. We highlight the most important ones as follows: 
\begin{itemize}
	\item \textbf{Skew / smile}: DLVs exhibit the \emph{volatility smile}, i.e. DLVs for deep in-the-money and out-of-the-money strikes are generally higher than at-the-money DLVs. \autoref{fig:dlv_smile} displays smile / smirk and shows the \emph{mean DLV} for each maturity for the Eurostoxx 50 dataset (see \autoref{fig:dlvseurostoxx}).
	\item \textbf{Persistence}: DLVs have high serial autocorrelation and the autocorrelation for short-dated DLVs drops generally faster than for long-dated ones. \autoref{fig:dlvs_atm_acf} displays the autocorrelation function for the first 128 lags for the ATM DLVs for maturities $20, 40, 60, 120$ business days; again for the Eurostoxx 50 dataset. 
	\item \textbf{Cross-correlations}: DLV strike-maturity pairs show high cross-correlation for proximate relative strikes and maturities. This is true for DLV levels as well as their returns (see \autoref{fig:dlvs_correlated}). 
\end{itemize} 

\section{Literature overview on market simulation}
\label{sec:literature_overview}
The absence of large equity option market datasets on a daily time scale has created attention from academia and industry. Although past works have studied modeling the dynamics and stylized facts of asset returns through diffusions in order to calibrate models for pricing under a risk-neutral measure $\mathbb{Q}$, they do not model the dynamics of the volatility surface under the real-world measure $\mathbb{P}$. Though traditional models can give an accurate fit of a single instance of the volatility surface they require constant day-to-day recalibration. 

Most models that fall into this category are parametric models that describe a family of distributions. In the financial mathematics literature examples are the Black-Scholes \cite{black2019pricing}, Heston \cite{heston1993closed}, SABR \cite{hagan2002managing} and rough volatility model \cite{gatheral2018volatility}. While these SDE-type models are explainable they face a trade-off between complexity and tractability which may cause difficulties in calibration. At the same time, the less complex a model becomes the higher the probability of model misspecification. In this paper, we use normalizing flows as a universal approximator for estimating the conditional discrete-time dynamics of the market \cite{durkan2019neural}. Our proposed market simulator, thus, is parametric in the sense that it describes a family of densities. However, through the high parametrisation of neural networks, our simulator is capable of universally approximating conditional densities such that the possibility of model misspecification is reduced\footnote{The generator formulation assumes a noise-state relationship. The incorporation of jumps is left as future work.}. 

More recently, with increasing interest in generative modeling, the focus has shifted on calibrating nonparametric models under the physical measure $\mathbb{P}$. \cite{arribas2020sigsdes} introduced Sig-SDEs, a signature-based SDE-type model, for pricing and risk management. \cite{Horvath2020} proposed to combine signatures with autoencoders to generate realistic spot dynamics but require an accurate projection from signature to path space. Very recently, \cite{cohen2021arbitrage} introduced an arbitrage-free neural SDE model which guarantees no dynamic arbitrage and minimizes the chance of generating static arbitrage. Various other works, including our single asset market simulator \cite{Wiese2019}, explored the application of GANs for simulating high-fidelity market data \cite{pardo2020mitigating, de2021tackling, koshiyama2019, Marti_2020, Hao2020, ni2021sigwasserstein, van2021monte, QuantGANs}.

Of the above mentioned works, our paper on simulating high-fidelity spot and equity option markets is most related to the work of \citeauthor{cohen2021arbitrage} \cite{cohen2021arbitrage} but differs in a number of aspects. First, our study is performed on real-world datasets, whereas \cite{cohen2021arbitrage} perform their study on a 2-factor Heston stochastic local volatility (SLV) model \cite{Jex1999PricingEU}. Second, while \citeauthor{cohen2021arbitrage} use a constrained PCA to find a linear call price basis for a lattice of call prices, we use an arbitrage-free parametrisation of a call price grid. This allows us to approximate the nonlinear pricing functional by the means of autoencoders \cite{hinton2006reducing} and enables us to learn efficient low-dimensional representations, whereas \cite{cohen2021arbitrage} need 3-factors to represent the ground truth 2-factor SLV model. Third, \cite{cohen2021arbitrage} enforce no dynamic arbitrage which is possible because they work under the risk-neutral measure $\mathbb{Q}$. We do not impose this constraint since we work under the real-world measure $\mathbb{P}$ which contains dynamic arbitrage. Last, our model is formulated in discrete-time whereas \citeauthor{cohen2021arbitrage} calibrate a continuous-time neural SDE. While a continuous-time simulator generating intraday tick prices and volumes may allow for enhanced risk management, we believe that this is a separate problem and leave it as future research.  

We emphasise that our approach of learning \emph{nonlinear and arbitrage-free} representations of the volatility surface by means of autoencoders and DLVs is novel and allows us to calibrate the pricing functional of the underlying. While several works have studied the application of PCA \cite{cohen2021arbitrage, cont2001empirical} and variational autoencoders \cite{Bergeron2021, Ning} for compression, these approaches are either limited by being linear, do not guarantee no static arbitrage, or cannot represent a rich family of call price surfaces. 

We highlight that we are not the first to apply normalizing flows to the application of modeling time series \cite{deng2021modeling} or conditional densities \cite{rothfuss2020noise}. \cite{deng2021modeling} proposed \emph{continuous-time flow processes} (CTFP) by leveraging neural ODEs for modeling the densities of the continuous-time process. While the CTFP solves the problem of sampling stationary time series on a fixed time horizon $[0, T]$ it does not solve the stationarity problem on a general time horizon $[0, T]$ where the terminal time $ T > 0$ is variable as the flow cannot generalise and extrapolate the non-stationary dynamics of the Brownian motion. Since downstream algorithms require sampling from the market simulator from variable terminal time $T \in (0, +\infty)$, we do not incorporate the continuous-time perspective. 

\section{Normalizing flows}
\label{sec:normalizing_flows_background}
\emph{Normalizing flows} \cite{dinh2016density,durkan2019neural, papamakarios2017masked, wehenkel2019unconstrained} are a class of functions within the function space of neural networks which are differentiable and bijective, more commonly referred to as \emph{diffeomorphisms}. In this section, we introduce the reader to the application of diffeomorphisms for modeling unconditional densities in \autoref{sec:unconditional}, discuss the existence of a diffeomorphism in \autoref{sec:existence}, how neural networks can be used to approximate bijective functions in \autoref{sec:normalizing_flows} and finally show how a network-based linear neural spline conditional on data can be constructed in \autoref{sec:linear_spline}.

\subsection{Modeling unconditional densities with normalizing flows}
\label{sec:unconditional}
Let $\bfz \sim \pz$ be a simple \emph{source density}, such as a multivariate normal, defined on a source space $Z \subseteq\R^d$, and $\bfx\sim\px^*$ a \emph{target density} defined on a compact target space $X \subseteq \R^d$. Assume that the target density is unknown. In generative modeling, the goal is to approximate a parametrisable map $T_\theta: \Z \to \X$ to transform the source distribution such that the model density $\bfx = T(\bfz) \sim \px$ and the target density $\bfx\sim\px^*$ are close with respect to some metric. 
Let $T_\theta: \Z \stackrel{\cong}{\to} \X$ be a diffeomorphism and denote by $\bfx=T(\bfz) \sim \px_\theta$ the model density induced by $T_\theta$. The density transformation theorem states that the density of $\bfx \sim \px_\theta$ admits the form
\begin{align}
	\label{eq:density_transform_1}
	\px_\theta(\bfx) &= \pz(\bfz) |\det J_{T_\theta}(\bfz)|^{-1}\\
	\label{eq:density_transform_2}
	&=\pz(T^{-1}_\theta(\bfx)) |\det J_{T_\theta}(T^{-1}_\theta(\bfx))|^{-1}
\end{align}
where $T^{-1}_\theta: \X \stackrel{\cong}{\to} \Z$ is the inverse, $\det$ denotes the matrix determinant operator and $J_{T_\theta}: \Z \to \mathbb{R}^{d \times d}$ denotes the Jacobian of $T_\theta$
\begin{equation*}
	J_{T_\theta}(\bfz) = 
	\left(
	\begin{array}{ccc}
		\dfrac{\partial T_1}{\partial \bfz_1} & \cdots & \dfrac{\partial T_1}{\partial \bfz_d}\\
		\vdots & \ddots & \vdots \\
		\dfrac{\partial T_d}{\partial \bfz_1} &\cdots & \dfrac{\partial T_d}{\partial \bfz_d}
	\end{array}
	\right)(\bfz)
\end{equation*}
where we dropped the dependence on $\theta$ for legibility. 

The tractability of the model density emerges as a great advantage as this property allows measuring the distance of the model and target density by employing $f$-divergences \cite{liese2006divergences} such as the Kullback-Leibler (KL) divergence. The KL-divergence of $\px^*$ and $\px_\theta$ is given as 
\begin{align}
	\KL(\px^*|| \px_\theta) &= \int_X \px^*(\bfx) \ln \dfrac{\px^*(\bfx)}{\px_\theta(\bfx)}d\bfx\\
	&= - \int_X \px^*(\bfx) \ln \dfrac{\px_\theta(\bfx)}{\px^*(\bfx)}d\bfx\\
	&= - \int_X \px^*(\bfx) \ln \dfrac{p(T^{-1}_\theta(\bfx))}{|\det J_{T_\theta}(T^{-1}_\theta(\bfx))|}d\bfx + \textrm{const}
	\label{eq:kl_div}
\end{align}
Hence, when $\px^*$ is available the KL-divergence can be evaluated numerically. However, in practice the target density is generally unknown and only a set of realizations $\lbrace \bfx_i \rbrace_{i=1}^n $ sampled from the target density $ \px^*$ is available. Thus, the KL divergence in \eqref{eq:kl_div} cannot be computed explicitly. Instead the integral is approximated via Monte Carlo: 
\begin{equation}
	\KL(\px^*|| \px_\theta) \stackrel{\textrm{MC}}{\approx} \dfrac{1}{n} \sum_{i=1}^n \left( \ln |\det J_{T_\theta}(T_\theta^{-1}(\bfx_i))| - \ln p(T^{-1}_\theta(\bfx_i)) \right) + \textrm{const}
	\label{eq:objective_unconditional}
\end{equation}

\begin{remark}[$f$-divergences]
	The KL-divergence is a single instance of a more broader class of $f$-divergences \cite{liese2006divergences}. The reason for the KL-divergence's popularity and application in this paper is that the diffeomorphism's parameters $\theta$ only depend on the negative expected log-model density under the target density
	\begin{equation*}
		\KL(\px^*|| \px_\theta) = - \mathbb{E}_{\bfx \sim \px^*}(\ln(\px_\theta(\bfx))) + \textrm{const}.
	\end{equation*}
	Note that this makes it possible, as opposed to other $f$-divergences, to approximate the $\theta$-dependant integral of the KL-divergence via Monte Carlo without having prior knowledge of the target density. 
\end{remark}

Normalizing flows approximate diffeomorphisms by applying gradient descent to the MC-approximated KL-divergence: 
\begin{align*}
	\nabla_\theta \KL(\px^* \| \px_\theta) &= - \mathbb{E}_{\bfx \sim \px^*}(\nabla_\theta\ln(\px_\theta(\bfx))) \\
	&\stackrel{\textrm{MC}}{\approx} 
	\dfrac{1}{n} \sum_{i=1}^n \nabla_\theta\left( \ln |\det J_{T_\theta}(T_\theta^{-1}(\bfx_i))| - \ln p(T^{-1}_\theta(\bfx_i)) \right)
\end{align*}

\subsection{Existence of a transport map}
\label{sec:existence}
In the previous section we assumed that diffeomorphisms can be used to transform any source density to some target density. In general, this only holds when the source and target density have overlapping supports. To prove the existence of a diffeomorphism that can transport a source density $\pz$ to a target density $\px$ we recall the notion of an \emph{increasing triangular map}. A function $T=(T_1, \dots, T_d): \R^d \to \R^d$ is called \emph{triangular} if for any $i = 1, \dots, d$ the function $T_i$ only depends on inputs $(\bfx_1, \dots, \bfx_i) \in \R^i$. Furthermore, we call a triangular map \emph{increasing} if for any $i=1, \dots, d$ the function $T_i: \R^d \to \R^d$ is monotonically increasing in $\bfx_i$. The following highly-celebrated result by \citeauthor{bogachev2005triangular} proved the existence of triangular maps for absolutely continuous Borel measures: 
\begin{thm}[\protect{\cite[Lemma 2.1]{bogachev2005triangular}}]
	\label{thm:bogachev}
	If $\mu$ and $\nu$ are absolutely continuous Borel probability measure on $\Z = \X = \R^d$, then there exists a unique (up to null sets of $\mu$) increasing triangular map $T : \Z \stackrel{\cong}{\to} \X$ such that $\nu = T_\#\mu$.\footnote{Here $T_\#\mu$ denotes the pushforward measure of $\mu$ which is defined for any Borel set $B \in \mathcal{B}(\X)$ as $T_\#\mu(B) = \mu(T^{-1}(B))$.} The same holds over $\Z = \X = [0,1]^d$. 
\end{thm}

The ability to approximate arbitrarily complex increasing triangular maps and diffeomorphisms is the core field of research of normalizing flows. 

\begin{remark}[Model misspecification]
	By constructing an algorithm that can approximate universally diffeomorphisms we leave the realm of \emph{quasi maximum likelihood} estimation, since for any absolutely continuous source / target density pair there exists a parametrised diffeomorphism $T_\theta$ that can transform the source to the target density. Thus, model misspecification is not possible anymore. 
\end{remark}

\subsection{Approximating diffeomorphisms with neural networks}
\label{sec:normalizing_flows}
In the previous two subsections, we assumed that the model density $q_\theta$ can be evaluated efficiently. However, this depends on the ability to compute the inverse $T^{-1}$ and the matrix determinant of the Jacobian $J_{T_\theta}$. In this subsection, we take a glance at the ideas that have been applied to approximate diffeomorphisms with neural networks such that both requirements on the map $T_\theta$ are fulfilled. We constrain ourselves to autoregressive flows and direct the reader to \cite{papamakarios2019normalizing} for a more in-depth overview on normalizing flows. 

Let $T_\theta: \Z \to \X$ be a diffeomorphism composed of $K$ diffeomorphisms
\begin{equation*}
	T_\theta = T_{K, \theta_K} \circ \cdots \circ T_{1, \theta_1}
\end{equation*}
and denote by $\bfz_{k} = T_{k, \theta_k} \circ \cdots \circ T_{1, \theta_1}(\bfz), k = 1, \dots, K$ the $k$-th successive composition / transformation and let $\theta = (\theta_1, \dots, \theta_K)$ denote the joint parameters. The log-determinant of the Jacobian of $T$ is given as 
\begin{equation*}
	\ln \det |J_T(\bfz)| = \ln \prod_{k = 1}^K |\det J_{T_{k}}(\bfz_{k-1})| = \sum_{k=1}^K \ln |\det J_{T_{k}}(\bfz_{k-1})|
\end{equation*}
where $\bfz_0 = \bfz$ and we dropped $\theta$-dependence for legibility. Hence, for the computation of the determinant to be efficient the computation of the determinants of $J_{T_k}, k=1,\dots, K$ needs to be efficient. To realize this requirement, it makes sense to construct the diffeomorphisms $T_k, k=1, \dots, K$ to be sufficiently \emph{simple}, but expressive enough to make $T$ universal. 

\subsection{Linear neural splines}
\label{sec:linear_spline}
In this subsection, we introduce linear neural splines \cite{durkan2019neural}. Linear neural splines are a special instance of normalizing flows and will be used in \autoref{sec:market_simulator} to approximate the market simulator. 

Neural splines follow the idea of approximating a diffeomorphism by constructing a triangular map. In the univariate case a neural spline degenerates to approximating the composition of the CDF of the source density and the inverse CDF of the target density.\footnote{Assume that the random variables $x \sim F$ and $y \sim G$ have CDFs $F, G: \R \stackrel{\cong}{\to} [0, 1]$ respectively. Then the optimal diffeomorphism transforming the source to the target distribution is defined as $H = G^{-1} \circ F: \R \stackrel{\cong}{\to} \R$.} Therefore, assume that the neural spline was to model a $D$-variate density $\bfx \sim \px^*$ defined on $[0, 1]^D$ by pushing forward an absolutely continuous source density $\bfz \sim \pz$. We also assume that we were to model the $i^{\textrm{th}}$ component and let $F_i: \mathbb{R}^{(i-1)} \times \Theta \to \R^N\times \mathbb{R}^N$ be a neural network with parameter space $\Theta$ and output space $\R^{N} \times \R^N$ where $N \in \N$ represents the \emph{number of knots} used to define the spline. Furthermore, denote by $\bfz_{<i} = (\bfz_1, \dots, \bfz_{i-1})$ the first $i-1$ components of the $D$-variate vector $\bfz \in \R^D$ and $(\tilde{\bfu}, \tilde{\mathbf{v}}) = F_i(\bfz_{<i})$ the output of the network. To construct the monotonically increasing spline the outputs $(\tilde{\bfu}, \tilde{\mathbf{v}})$ need to be transformed such that they represent monotonically increasing coordinates in $[0, 1]^2$. For this purpose, let $\operatorname{softmax}: \R^N \to [0, 1]^N$ be the softmax function and $\operatorname{cumsum}: \R^N \to \R^N$ the cumulative sum function which computes the successive cumulative sum. The spline's $[0, 1]^2$-valued coordinates can be computed by applying the softmax and the cumulative sum function
\begin{equation*}
	\bfu = \operatorname{cumsum}(\operatorname{softmax}(\tilde{\bfu})), \quad \mathbf{v} =  \operatorname{cumsum}(\operatorname{softmax}(\tilde{\mathbf{v}})).
\end{equation*}
Finally, we are able to define the neural spline $T: \R^{i-1} \times \Theta \to (\mathbb{R} \stackrel{\cong}{\to} [0, 1] ) $ as 
\begin{equation*}
	T_i(\bfz_i; \bfz_{i<}, \theta) \coloneqq  
	\left\lbrace
	\begin{array}{cc}
		\mathbf{v}_j + (\bfz_i-\bfu_j) \dfrac{(\mathbf{v}_{j+1} - \mathbf{v}_j)}{(\bfu_{j+1} - \bfu_j)} & \quad \textrm{if} \ \bfz_i \in [\bfu_j, \bfu_{j+1}] 
	\end{array}
	\right. 
\end{equation*}

Above we assumed that target density had support on the hypercube $[0, 1]^D$. Neural splines can be generalized to $\R^D$ by applying an affine transformation to the coordinates and extrapolating the tail behaviour through predefined / extrapolated tails. Furthermore, generalisations of the here introduced linear spline construction have been introduced in \cite{durkan2019neural} for monotonic rational-quadratic splines and in \cite{durkan2019cubic} for cubic splines. 

\section{Simulating single-asset spot and equity option markets with normalizing flows}
\label{sec:market_simulator}
After introducing spot and equity option markets and normalizing flows, we formally present in this section the proposed flow-based market simulator for a single underlying. The section is divided into four parts: 
\begin{itemize}
	\item Subsection \ref{sec:mono_compression}: nonlinear arbitrage-free compression of the call price surface,
	\item Subsection \ref{sec:mono_calibration}: derivation of the market simulator objective, 
	\item Subsection \ref{sec:mono_martingale}: modifications to the simulator to guarantee the martingale property, 
	\item Subsection \ref{sec:mono_numerical}: numerical results demonstrating the performance of the calibrated Eurostoxx 50 market simulator.
\end{itemize}
Throughout this section, we denote by $I \coloneqq \lbrace 1, \dots, T\rbrace$ the time horizon, for $t \in I $ by $\boldsymbol{\Sigma}_{t} \in \mathbb{R}^{mn}_{>0}$ and $s_t$ the historical $m \times n$-dimensional grid of DLVs and the spot level respectively and for $t \in I \setminus \lbrace 0 \rbrace $ by $r_t = \ln(s_t) - \ln(s_{t-1})$ the spot log-return.  

\subsection{Learning efficient arbitrage-free representations of the call price surface}
\label{sec:mono_compression}
DLV levels as well as their returns are highly correlated for proximate relative strikes and maturities (see \autoref{fig:dlvs_correlated}). Since our aim is to calibrate a market simulator from limited real-world market data it does not make sense to the full joint distribution of each DLV strike-maturity combination due to the curse of dimensionality \cite{bishop2006pattern}. We therefore reduce the dimensionality of the grid of DLVs using an autoencoder. Autoencoders are neural networks with a \emph{bottleneck structure} that are trained to learn the identity mapping through a reconstruction error. The bottleneck structure forces the network to learn a low-dimensional encoding of the DLV grid. An autoencoder can be split up into two parts: an encoder function $E:\mathbb{R}_{}^{mn} \to \mathbb{R}^{d-1}$ which compresses the grid and a decoder function $D: \mathbb{R}^{d-1} \to \mathbb{R}^{mn}_{}$ reconstructing / decompressing the encoding. 

The only requirement that needs to be satisfied in order to ensure no static arbitrage in the reconstructed call prices which we obtain from inverting the reconstructed DLVs (see \autoref{sec:no_static_arb}) is that the reconstructed DLVs are non-negative. We therefore preprocess DLVs to log-DLVs to ensure that the DLVs are non-negative and then standard scale them
\begin{align*}
	\boldsymbol{\Sigma}^{\ln}_t &= \ln(\boldsymbol{\Sigma}_t)\\
	\boldsymbol{\alpha}_t &= (\gamma_t - \hat{\mu}_{\boldsymbol{\Sigma}^{\ln}}) \oslash \hat{\gamma}_{\boldsymbol{\Sigma}^{\ln}}
\end{align*}
where $\hat{\mu}_{\boldsymbol{\Sigma}^{\ln}}$ and $\hat{\gamma}_{\boldsymbol{\Sigma}^{\ln}}$ is the component-wise empirical mean and standard deviation respectively and $\oslash$ denotes component-wise division. 

To demonstrate the efficacy of our DLV autoencoder approach we compare the autoencoder (\emph{AE}) with a \emph{principal component analysis} (\emph{PCA}) (see \cite[Chapter 12.1]{bishop2006pattern}) performed on the DLVs. We shuffle the preprocessed data $(\boldsymbol{\alpha}_t)_{t \in I}$ via random permutations and then split the shuffled data into a train and test dataset $I^{\textrm{train}} \cup I^{\textrm{test}} = I$; where $I^{\textrm{train}}$ denotes the indices of the training set and $I^{\textrm{test}}$ the indices of the test. We retain $80\%$ of the data within the train dataset; $20\%$ are used in the test set. The autoencoder is implemented using the high-performance deep learning library \emph{pytorch}  \cite{paszke2017automatic, pytorch2019} and is trained using the Adam optimizer \cite{kingma2014adam} with a learning rate of $0.001$ and $2\%$ DropOut \cite{srivastava14a} across all layers to mitigate overfitting. The loss function used is the \emph{mean-squared error} (\emph{MSE}), also known as the \emph{image loss} or \emph{reconstruction error};
\begin{equation*}
	\min_{(\theta_1, \theta_2) \in \Theta} (|I^{\textrm{train}}|mn)^{-1}\sum_{t \in I^{\textrm{train}}} \left\|\boldsymbol{\alpha}_t - D_{\theta_2}(E_{\theta_1}(\boldsymbol{\alpha}_t))\right\|_2^2  
\end{equation*}  
where $\|\cdot\|_2$ denotes the $L_2$-norm. Note that we scale the sum by the scalar constant $(|I^{\textrm{train}}|mn)^{-1}$ so that the learning rate is invariant to a change in the size of the dataset or the DLV grid.

\begin{figure}[!h]
	\centering
	\begin{minipage}{.5\textwidth}
		\centering
		\includegraphics[width=\textwidth]{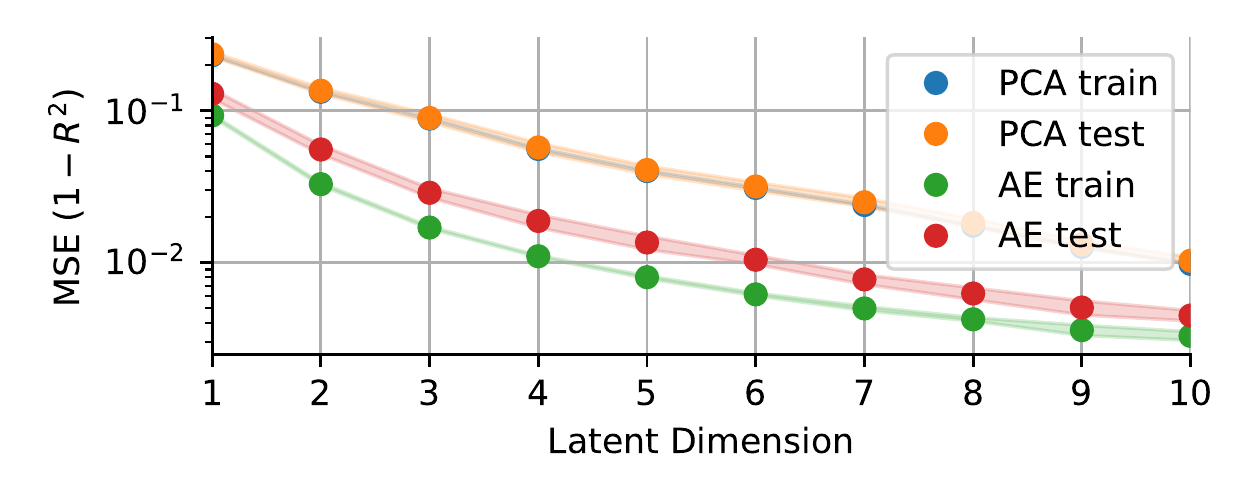}
		\caption{Reconstruction error (MSE) obtained by calibrating autoencoders / principal components for the considered latent dimensions.}
		\label{fig:ae_pca_mse}
	\end{minipage}%
	\begin{minipage}{.5\textwidth}
		\centering
		\includegraphics[width=\textwidth]{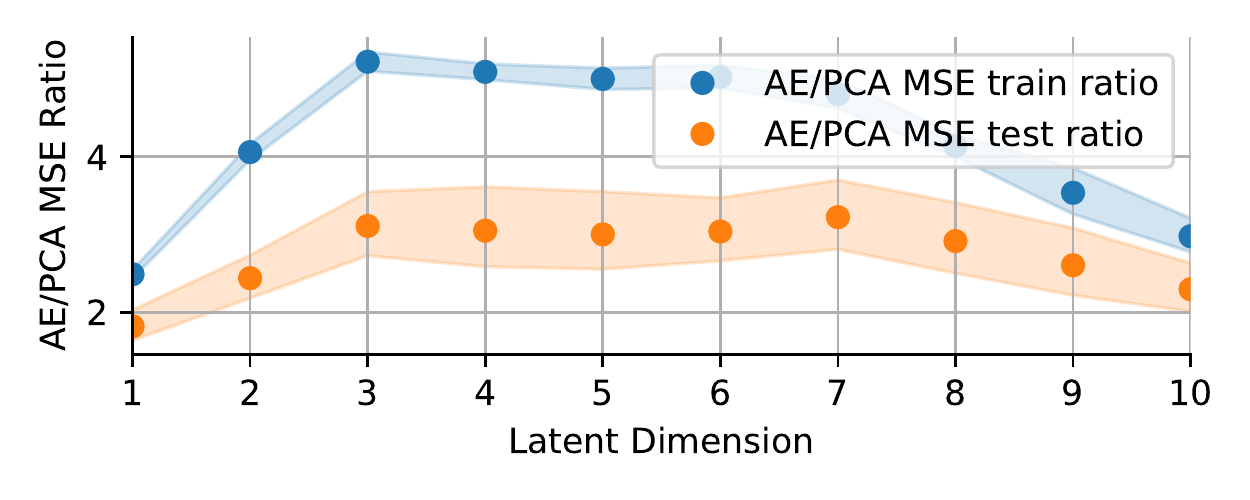}
		\caption{AE / PCA reconstruction error ratio.\\~\\~}
		\label{fig:ae_pca_mse_ratio}
	\end{minipage}
\end{figure}

\begin{figure}[htp]
	\centering
	\includegraphics[width=\textwidth]{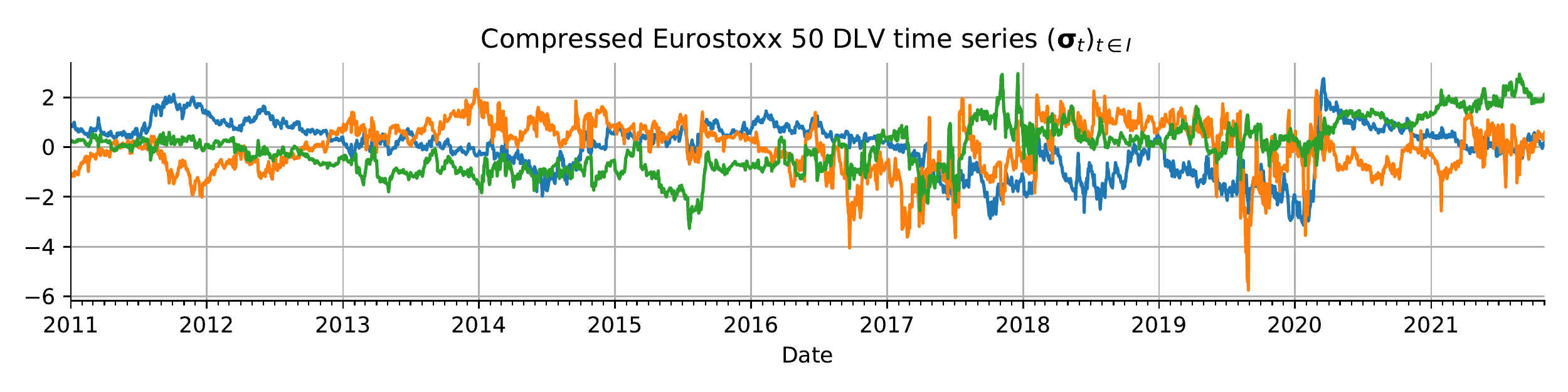}
	\caption{Three-dimensional compressed representation of the Eurostoxx 50 grid obtained through a calibrated nonlinear encoder function. While the blue encoding exhibits properties of the overall volatility level, the orange and green time series encode features such as the term structure and volatility skew.}
	\label{fig:compressed_components}
\end{figure}

The above training procedure is performed across ten different seeds and latent dimensions one through ten. For each seed and latent dimension we store the MSE on the train and test set. The \emph{mean MSE} obtained across all seeds for each individual latent dimension is reported in \autoref{tab:mse_autoencoder} and visualised additionally in \autoref{fig:ae_pca_mse}. The error bars around each mean MSE show the standard deviation of the MSEs obtained for each latent dimension to give a sense of the variation and reproducibility we obtained during training. Furthermore, we display the AE and PCA MSE ratio in \autoref{fig:ae_pca_mse_ratio} which shows that in terms of the MSE the AE beats PCA by a factor of two to five on the train set and two to three on the test set highlighting the efficiency of the DLV-AE approach in contrast to a linear encoding. \autoref{fig:compressed_components} displays a three-dimensional encoding of the DLV grid obtained through a trained autoencoder. 

\begin{table}[!h]
	\resizebox{\textwidth}{!}{
	\begin{tabular}{lcccccccccc}
		Latent dimension & 1 & 2 & 3 & 4 & 5 & 6 & 7 & 8 & 9 & 10\\ \toprule
		AE train set & $0.093$ & $0.033$ & $0.017$ & $0.011$ & $0.008$ & $0.006$ & $0.005$ & $0.004$ & $0.004$ & $0.003$\\
		PCA train set & $0.232$ & $0.133$ & $0.089$ & $0.056$ & $0.040$ & $0.031$ & $0.024$ & $0.017$ & $0.013$ & $0.010$\\
		\toprule
		AE test set & $0.130$ & $0.056$ & $0.029$ & $0.019$ & $0.014$ & $0.010$ & $0.008$ & $0.006$ & $0.005$ & $0.004$\\
		PCA test set & $0.236$ & $0.136$ & $0.090$ & $0.057$ & $0.041$ & $0.032$ & $0.025$ & $0.018$ & $0.013$ & $0.010$
	\end{tabular}
	}
	\caption{MSE $(1- R^2)$ of the reconstructed standard-scaled log-DLVs surfaces $(\boldsymbol{\alpha}_t)_{t \in I}$ obtained through PCA and autoencoding for bottleneck dimensions one through ten.}
	\label{tab:mse_autoencoder}
\end{table}

\subsection{Derivation of the market simulator's objective}
\label{sec:mono_calibration}
In the last section, we addressed the curse of dimensionality by the means of a DLV-based autoencoder which is able to learn an efficient low-dimensional encoding of the call price surface while satisfying no static arbitrage in the reconstructed surface. In this section, we construct the market simulator and derive the objective function. 

Assume that there is a calibrated encoder function $E:\mathbb{R}^{nm} \to \mathbb{R}^{d-1}$ that encodes the preprocessed DLV grid $(\boldsymbol{\alpha}_t)_{t \in I}$ to the compressed space and denote the compressed process by $\boldsymbol{\sigma}_t = E(\boldsymbol{\alpha}_t), t \in I$. Furthermore, denote by $\bfx_t = (r_t, \boldsymbol{\sigma}_t)$ for $t \in T$ the \emph{market state}. Our aim in this section is to calibrate a model density $\px_\theta$ such that for any $p$-lagged market states $\bfy_t = (\bfx_t, \dots, \bfx_{t-p+1}) \in \Y=\X^p$ and next-day market state $\bfx_{t+1} \in X$ the target and model conditional densities $\px^*(\bfx_{t+1} | \bfy_t)$ and $\px_\theta(\bfx_{t+1} | \bfy_t)$ are close with respect to the Kullback-Leibler divergence. 

Here, the conditional model density $\px_\theta(\bfx_{t+1} | \bfy_t)$ will be represented through a conditional normalizing flow where the flow will not only be parametrised by some parameter space $\Theta$ but also by the past market states $\bfy_t\in\Y$. Let 
\begin{align*}
	T_\theta: \Y &\to (\Z \stackrel{\cong}{\to} \X) \\
	\bfy_t &\mapsto T(\cdot; \bfy_t)
\end{align*}
be such a function mapping the $p$-lagged states $\bfy_t\in\Y$ to a diffeomorphism, i.e. for fixed $\bfy_t \in \Y$ the function $T_\theta(\cdot; \bfy_t): \Z \stackrel{\cong}{\to} \X$ is a diffeomorphism. Likewise to the unconditional case described in \autoref{sec:unconditional} the model density conditional on $\bfy_{t} \in \Y $ can be expressed via the density transformation theorem
\begin{align}
	\px_\theta(\bfx_{t+1}; \bfy_t)=p(T^{-1}_\theta(\bfx_{t+1}; \bfy_{t})) |\det J_{T_\theta}(T^{-1}_\theta(\bfx_{t+1}; \bfy_{t}); \bfy_{t})|^{-1}
\end{align}

For calibration we use the \emph{expected conditional KL-divergence} since our interest is to calibrate the conditional density not only for a single  condition  $\bfy \in Y$ but across all possible past conditions weighted by their likelihood
\begin{align*}
	\mathcal{L}(q^*, q_\theta) &\coloneqq \mathbb{E}_{\bfy\sim\px^*}\left(\KL\left(\px^*(\cdot | \bfy_t)\|\px_\theta(\cdot | \bfy_t)\right) | \bfy_t = \bfy\right) \\
	&=-\int_Y\int_X \px^*(\bfx_{t+1} | \bfy_t)\px^*(\bfy_t) \ln\dfrac{\px_\theta(\bfx_{t+1} | \bfy_t)}{\px^*(\bfx_{t+1} | \bfy_t)} d\bfx_{t+1} d\bfy_t\\
	&= - \int_Y \int_X \px^*(\bfx_{t+1}, \bfy_t) \ln \dfrac{p(T^{-1}_\theta(\bfx_{t+1}; \bfy_t))}{|\det J_{T_\theta}((T^{-1}_\theta(\bfx_{t+1}; \bfy_t); \bfy_t)|} d\bfx_{t+1} d\bfy_t + \textrm{const}
\end{align*}
In practice, due to the limited amount of market data, the ground truth density $\px^*$ is not available. Instead only a realization $(\bfx_t)_{t \in I}$ exists. We therefore approximate the expected conditional KL-divergence via Monte Carlo: 
\begin{equation}
	\label{eq:nll_conditional}
	\mathcal{L}^{\textrm{MC}}(q^*, q_\theta) := \dfrac{1}{T-p} \sum_{t=p}^{T-1} \ln |\det J_{T_\theta}(T^{-1}_\theta(\bfx_{t+1}; \bfy_t); \bfy_t)| - \ln p(T^{-1}_\theta(\bfx_{t+1}; \bfy_t)) 
\end{equation}
where we dropped the constant term for legibility. Minimization of the objective is performed via gradient descent. 

\begin{figure}[htp]
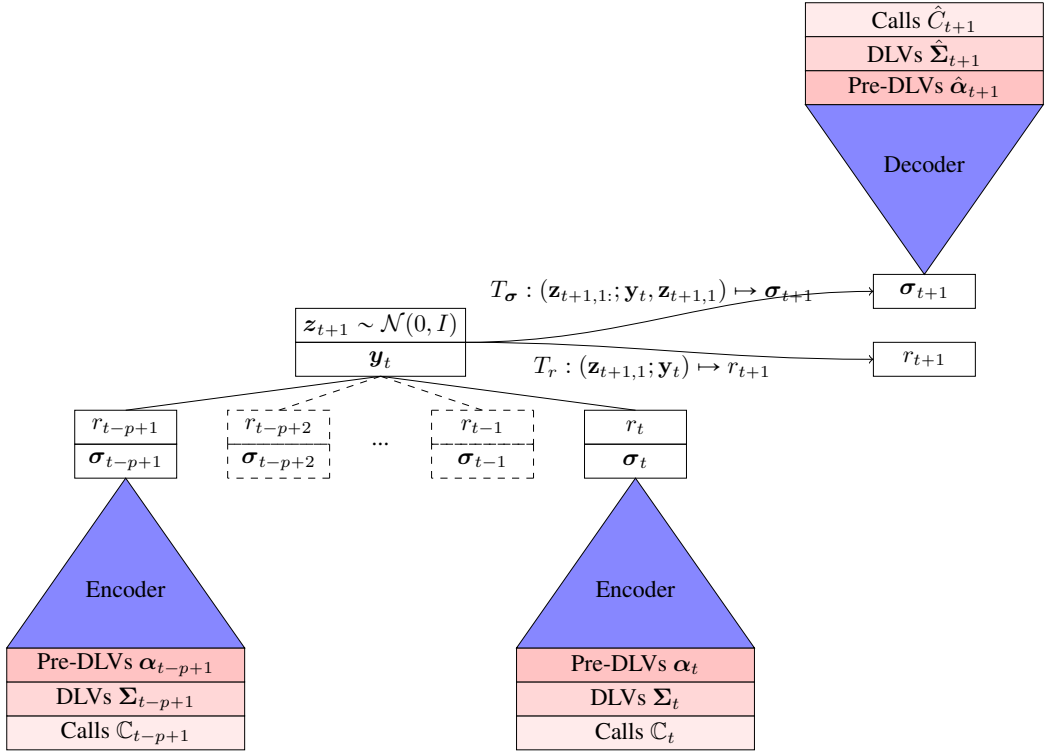

    \ctikzfig{simulator}
    \caption{Schematic diagram of the conditional spot and equity option market simulator $T_{r} $ and $T_{\boldsymbol{\sigma}}$ respectively.}
    \label{fig:conditionalsimulator}
\end{figure}

\subsection{Addressing non-zero drift in generated spot prices}
\label{sec:mono_martingale}
Conditional drift in the generated spot prices is the most obvious form of statistical arbitrage which can be exploited by a downstream algorithm such as Deep Hedging. If the spot drift conditional on $\bfy_t \in \Y$ is positive (negative) Deep Hedging will buy (sell) the underlying and make a profit in expectation. From daily spot data it’s hard to predict non-zero drift with any confidence. In our application, we do not want to be sensitive to drift. We therefore choose to ensure the martingale property when generating spot prices. 

In order to satisfy the martingale property, we impose a normal distribution onto the generated log-returns
\begin{equation}
	\label{eq:normal_form}
	r_{t+1} \sim \mathcal{N}\left(\frac{-\nu^2_t}{2}, \nu^2_t\right)
\end{equation}
where $\nu_t = F_\nu(\bfy_t)$ denotes the spot market’s realized volatility and the map $F_\nu: \Y \to \mathbb{R}_{>0}$ the realized volatility estimator; in our application a neural network. While we do not claim that the historical spot log-returns follow a normal distribution, the proposed constraint is a first step to generating realistic martingale dynamics containing e.g. stochastic volatility, and we leave it as future work to relax the constraint \eqref{eq:normal_form}. \autoref{fig:conditionalsimulator} illustrates the construction and design of the spot and equity option simulator $(T_r, T_{\boldsymbol{\sigma}})$ where the transport map $T_r$ is defined as
\begin{equation*}
	T_r: (\bfz_{t+1}, \bfy_t) \mapsto \bfz_{t+1, 1} F_\nu(\bfy_{t}) - \dfrac{F_\nu(\bfy_{t})^2}{2}
\end{equation*} 
for $\bfz_{t+1} \sim \mathcal{N}(0, 1)$. 



\subsection{Numerical results: Eurostoxx 50}
\label{sec:mono_numerical}
In this section, we present numerical results and demonstrate empirical evidence that the trained market simulator performs well on a real-world single asset dataset. We consider the Eurostoxx 50 from January 2011 to November 2021 which amount to a total of 2711 business days. The DLV grid will be defined for relative strikes $\lbrace 0.8, 0.85, \dots, 1.2\rbrace$ and maturities $\lbrace 20, 40, 60, 120 \rbrace$; here quoted in business days. \autoref{fig:dlvseurostoxx} displays the considered DLV time series. 

Prior to calibrating the neural spline an autoencoder is trained as described in \autoref{sec:mono_compression} using a three-dimensional bottleneck. \autoref{fig:compressed_components} depicts the compressed state representation. Afterwards, the compressed representation is standard-scaled (z-transformed) and concatenated to the spot log-return series to obtain the market state time series $(\bfx_t)_{t \in I}$; where $I$ is the time set. Next, a rolling-window is applied to obtain the rolled time series $ (\bfy_{t}, \bfx_{t+1})_{t=p}^{T-1}$ containing the $p$-lagged market states and the \emph{next-day} market state. We chose $p=2$ for the lagged window as we observed that the second lag of the partial autocorrelation function of the compressed DLVs is non-neglibly far away from zero. The rolled series is then split into a train and test set via random permutations where $80\%$ of the data are held in the train set; the rest is retained in the test set. 

\begin{figure}[!htb]
    \centering
    \begin{minipage}{.32\textwidth}
        \centering
        \includegraphics[width=\textwidth]{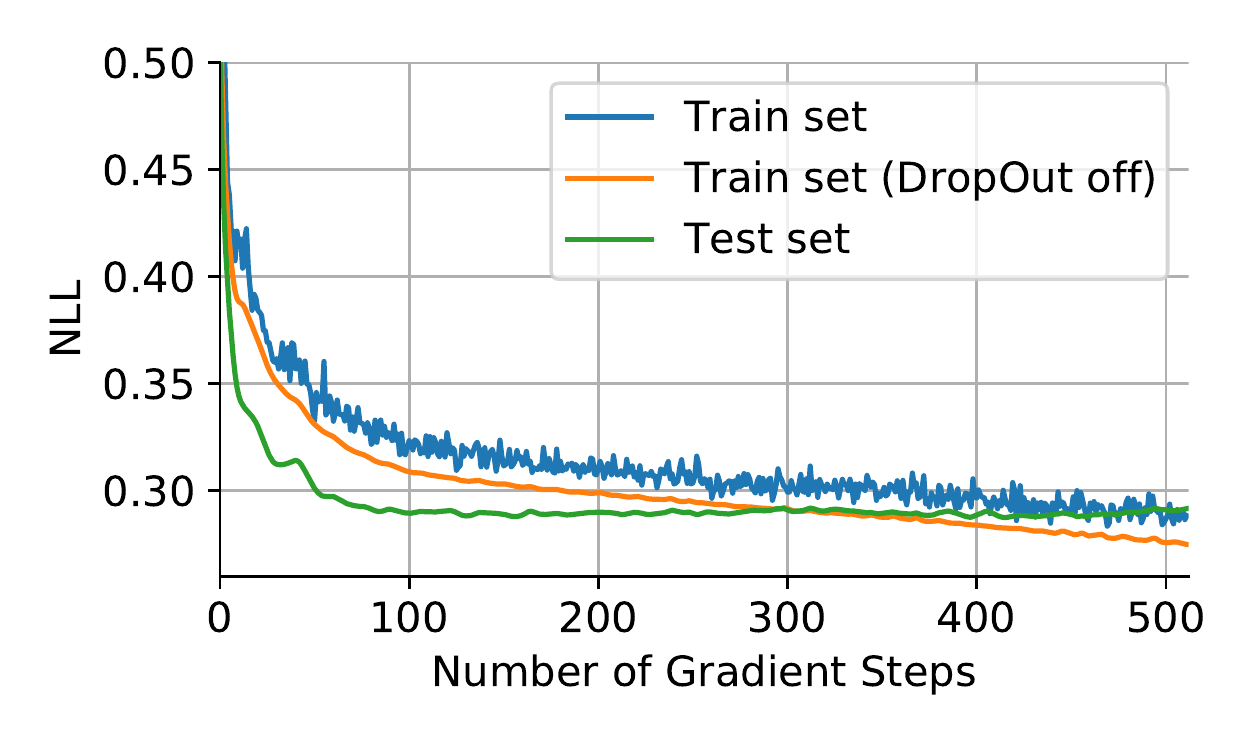}
		\caption{Development of train and test set errors obtained during optimization of $T_r$.}
        \label{fig:t_nu_losses}
	\end{minipage}%
	\,
    \begin{minipage}{.32\textwidth}
        \centering
		\includegraphics[width=\textwidth]{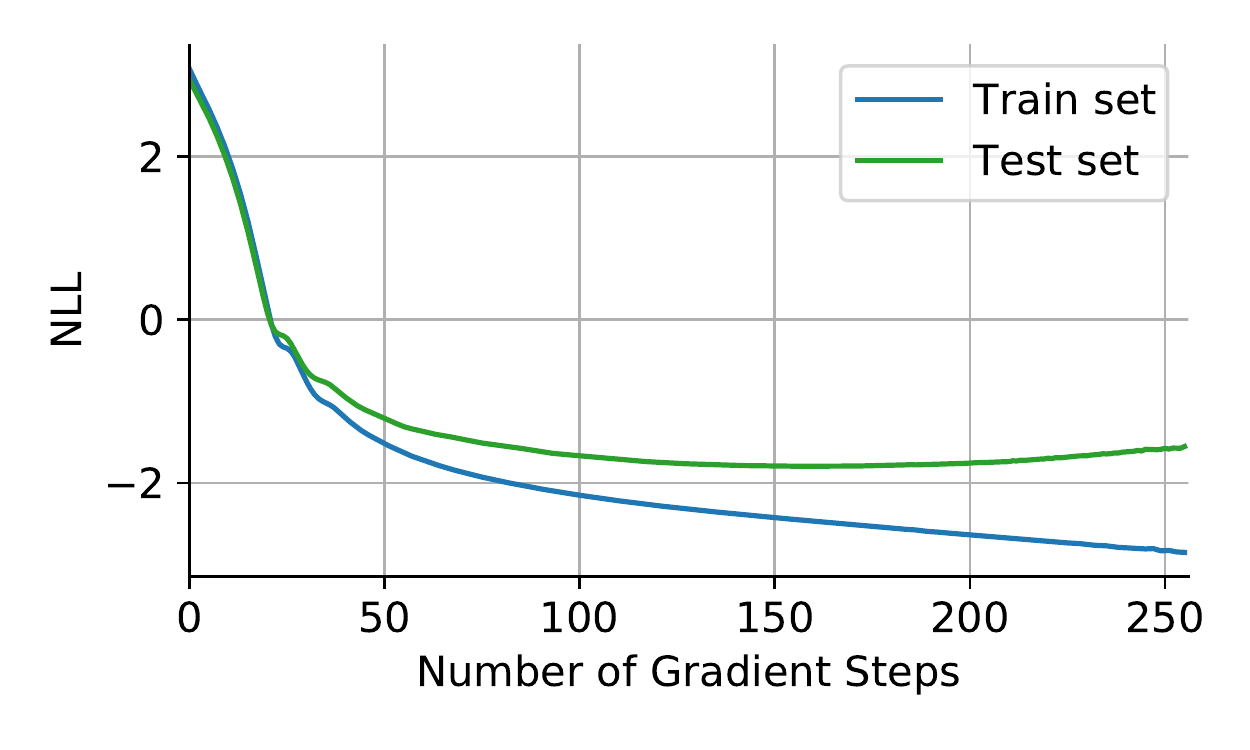}
		\caption{Development of train and test set error obtained during optimization of $T_{\boldsymbol{\sigma}}$.}
        \label{fig:t_sigma_losses}
	\end{minipage}
	\,
	\begin{minipage}{.32\textwidth}
        \centering
		\includegraphics[width=\textwidth]{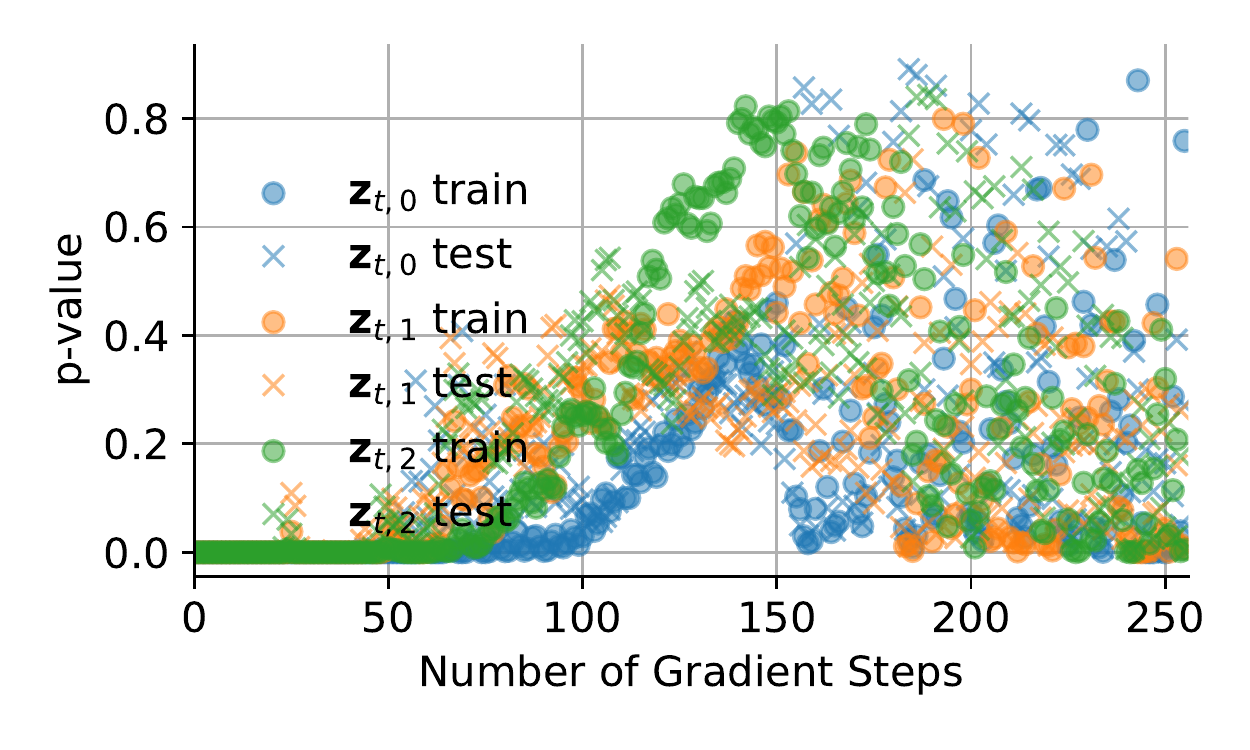}
		\caption{Development of the KS $p$-value during calibration of $T_r$.\\~}
        \label{fig:ks_p_value}
    \end{minipage}
\end{figure}

After compressing the DLVs and preparing the data the realized volatility estimator $T_r$ is trained to maximize the likelihood under the assumption that the spot log-return distribution follows a conditionally normal form \eqref{eq:normal_form}. To approximate the realised volatility estimator $F_\nu$ we use a three layer neural network with 64 hidden dimensions and apply the exponential function as a last transformation to ensure that the predicted realized volatility is positive. The network is trained with the Adam optimizer, a learning rate of $0.001$ and $10\%$ DropOut to address overfitting. \autoref{fig:t_nu_losses} shows the decay of the loss function during the course of training. Early stopping is applied at the $300^{th}$ gradient step / iteration to stop the network from overfitting as at this point the test and train loss (when DropOut is turned off) intersect. 

Next, the compressed state simulator map is calibrated through the NLL objective \eqref{eq:nll_conditional} and batch gradient descent. The flow is represented through three neural networks, each constructing a monotonically increasing linearly interpolated spline which will be used to transform the sampled Gaussians $\bfz_{t, 1:} \sim \mathcal{N}(0, I)$ to the next-day compressed DLVs $\boldsymbol{\sigma}_{t+1}$. Each network-based spline approximator will have three hidden layers with $64$ hidden dimensions. Furthermore, we use the Adam optimizer with a learning rate of $0.001$ for calibration. $80\%$ of the data is kept in the train set and the rest is used in the test set to assess whether the model is overfitting. \autoref{fig:t_sigma_losses} shows the development of the NLL on the train (blue) and test set (green). We observed that the test set error stopped to decay at the $150^{th}$ iteration and applied early stopping there.  

\begin{figure}[htp]
	\centering
	\includegraphics[width=\textwidth]{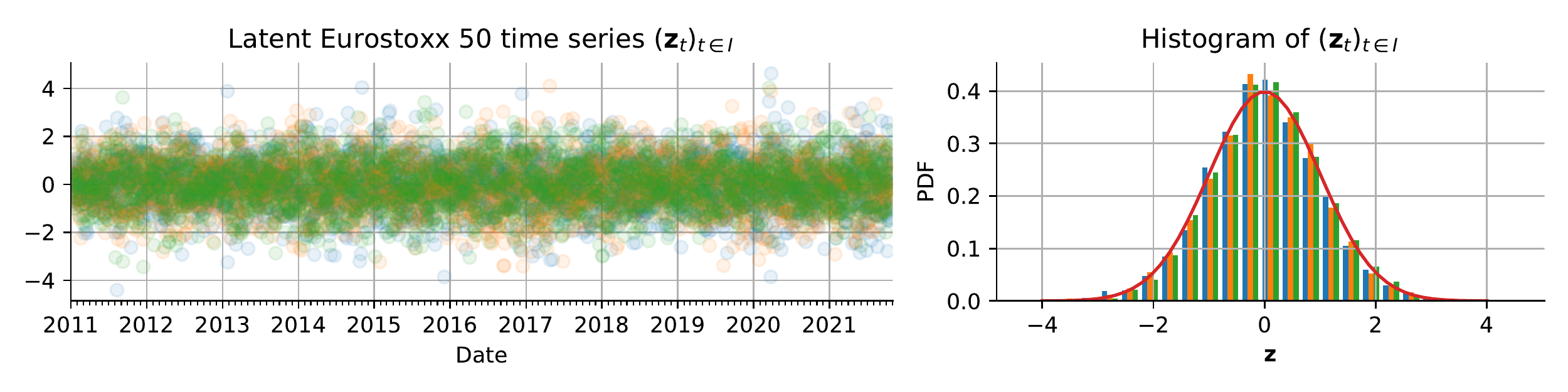}
	\caption{Latent process of Eurostoxx 50 and marginal densities.}
	\label{fig:noise_marginals_sx5e}
\end{figure}

After training the full market simulator $T = (T_r, T_{\boldsymbol{\sigma}})$ we measure the goodness of the fitted model. Evaluation is split into two parts: assessment of the latent time series and performance on distributional and dependency measures. The latent time series is computed by inverting the rolled time series $\bfz_{t+1} = T^{-1}(\bfx_{t+1}, \bfy_t), t \in I \setminus \lbrace 1, \dots, p-1, T \rbrace $. \autoref{fig:noise_marginals_sx5e} depicts the latent series as well as its empirical distribution. Visually, we can observe that the latent series resembles properties of a normally distributed i.i.d process. Furthermore, we can verify that the empirical probability density function (PDF) is close to the standard normal PDF (displayed in red in \autoref{fig:noise_marginals_sx5e}). To further assess the proximity of the latent unconditional distribution to the standard normal we apply a Kolmogorov-Smirnov (KS) test \cite{massey1951kolmogorov} to each component individually and test whether we can reject the null hypothesis ($H_0$) that the samples were sampled from a standard normal distribution. \autoref{tab:ks_statistics} reports the $p$-values obtained from the KS test and we conclude that we cannot reject $H_0$ at a (approximately) $72\%$ and $75\%$ confidence level on the train and test set respectively. 

\begin{table}[]
	\centering
	\begin{tabular}{@{}lcccccc@{}}
		Data set              & \multicolumn{3}{c}{Train set} & \multicolumn{3}{c}{Test set} \\ \midrule
		Latent component & 1        & 2        & 3       & 1        & 2       & 3       \\ \midrule
		KS $p$-value     & 0.2889   & 0.5624   & 0.7962  & 0.3335   & 0.2586  & 0.2898
	\end{tabular}
\caption{Kolmogorov-Smirnov $p$-values of the latent components on the train and test set.}
\label{tab:ks_statistics}
\end{table}
During calibration we additionally tracked the $p$-values on the train and test set depicted in \autoref{fig:ks_p_value}. Interestingly, the $p$-values increase on the train and test set until the iteration $140-160$ and then start to rapidly decrease which coincides with the iteration range where the test set error starts to increase. 

Our second assesment is based on the distributional and dependence properties on a short and large time horizon. We assess these properties individually on a varying time horizon as we want to determine whether short-term and the long-term \emph{invariant} distribution of the simulator is proximate to the distribution of the data. For evaluation on the short time horizon we sample for each condition $\bfy_t, t \in I \setminus \lbrace 1, \dots p-1 \rbrace$ a total of $M=4$ of length $\tau=3$
\begin{equation*}
	\lbrace\bfx_{t+\tau}^{(i)}, \dots, \bfx_{t+1}^{(i)}\rbrace_{i=1}^M \sim \px_\theta(\cdot | \bfy_t)  
\end{equation*}
resulting in $M \cdot (T-p)$ paths of length $\tau$. We then assess the distributional and dependence properties of the level process $(r_t, \boldsymbol{\sigma}_t)$ in \autoref{fig:sx5e_generative_performance_short_term_level} and return process $(r_t, \Delta\boldsymbol{\sigma}_t)$, where $\Delta \boldsymbol{\sigma}_t$ is denotes the return of the compressed DLV grid, i.e. $\Delta \boldsymbol{\sigma}_t = \boldsymbol{\sigma}_{t} - \boldsymbol{\sigma}_{t-1}$, in \autoref{fig:sx5e_generative_performance_short_term_return}. Figures \ref{fig:sx5e_generative_performance_short_term_level} and \ref{fig:sx5e_generative_performance_short_term_return} show from left to right histograms of the unconditional historical (blue) and generated (orange) distribution, as well as the autocorrelation function for the first $2$ lags and the historical and generated cross-correlation matrix. Visually, we can verify that the NLL objective used to calibrate the market simulator leads to a market simulator that approximates these statistical properties. 

\begin{figure}[h!]
	\centering
	\includegraphics[width=\textwidth]{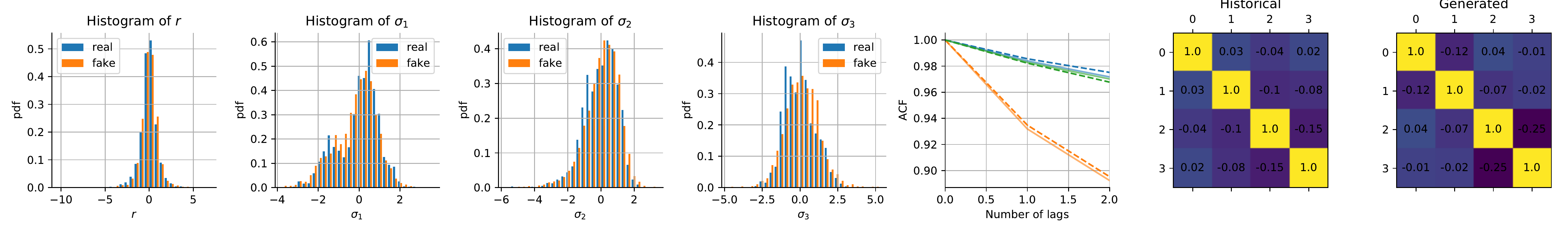}
	\caption{Eurostoxx 50 long-term performance of the level process $(r_t, \boldsymbol{\sigma}_t)$. From left to right: histograms of $r, \boldsymbol{\sigma}_{t, 1}, \boldsymbol{\sigma}_{t, 2}, \boldsymbol{\sigma}_{t, 3}$ where blue indicates the color of the historical dataset and orange the color of the generated, ACF of the level process for the first two lags, and last the cross-correlation matrix of the historical (left) and generated (right).}
	\label{fig:sx5e_generative_performance_long_term_level} 
\end{figure}

\begin{figure}[h!]
	\centering
	\includegraphics[width=\textwidth]{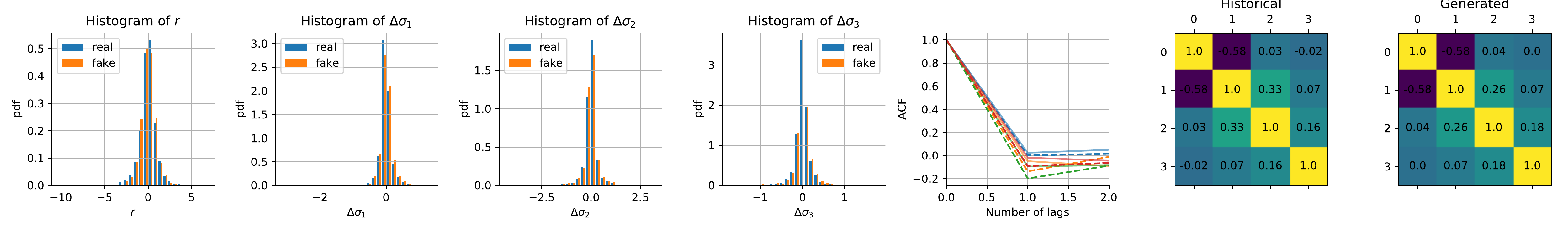}
	\caption{Eurostoxx 50 long-term performance of the return process $(r_t, \Delta \boldsymbol{\sigma}_t)$. From left to right: histograms of $r_t, \Delta \boldsymbol{\sigma}_{t, 1}, \Delta \boldsymbol{\sigma}_{t, 2}, \Delta \boldsymbol{\sigma}_{t, 3}$ where blue indicates the color of the historical dataset and orange the color of the generated, ACF of the return process for the first two lags, and last the cross-correlation matrix of the historical (left) and generated (right).}
	\label{fig:sx5e_generative_performance_long_term_return} 
\end{figure}

\begin{figure}[h!]
	\centering
	\includegraphics[width=\textwidth]{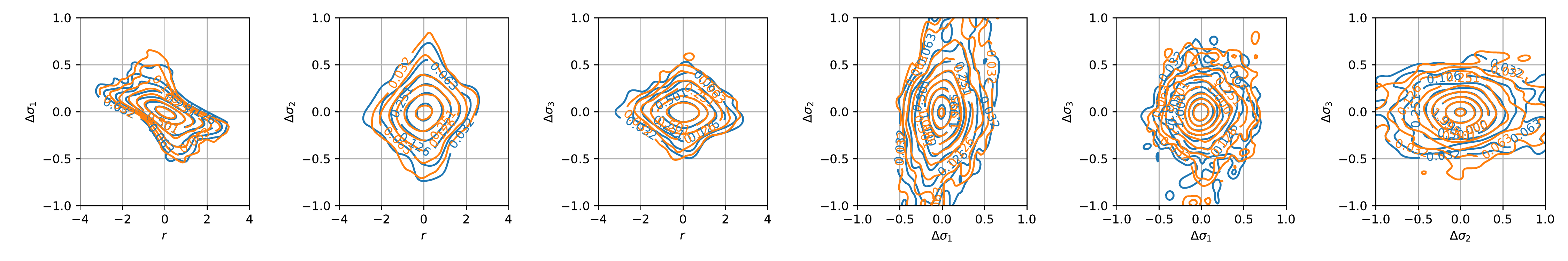}
	\caption{Eurostoxx 50 kernel density estimate of the joint return distribution $(r_t, \Delta \boldsymbol{\sigma}_t)$ of the empirical (blue) and generated distribution (orange). The kernel density estimates show from left to right the joint distribution of $(r_t, \Delta \boldsymbol{\sigma}_{t, 1}), (r_t, \Delta \boldsymbol{\sigma}_{t, 2}), (r_t, \Delta \boldsymbol{\sigma}_{t, 3}), (\Delta\boldsymbol{\sigma}_{t, 1}, \Delta \boldsymbol{\sigma}_{t, 2}), (\Delta\boldsymbol{\sigma}_{t, 1}, \Delta \boldsymbol{\sigma}_{t, 3}) $ and $ (\Delta\boldsymbol{\sigma}_{t, 2}, \Delta \boldsymbol{\sigma}_{t, 3})$.}
	\label{fig:sx5e_kde_return}
\end{figure}

The same procedure as above is performed for evaluation of the long-term properties, however, here we sample for $\tau=256$ days and only retain the last $4$ days
\begin{equation*}
	\lbrace\bfx_{t+\tau}^{(i)}, \dots, \bfx_{t+\tau - 3}^{(i)}\rbrace_{i=1}^M \sim \px_\theta(\cdot | \bfy_t), t \in I \setminus \lbrace 1, \dots p-1 \rbrace.
\end{equation*}
During simulation of long-term paths we noticed that approximately $2.5\%$ of the paths exploded and had to be discarded / rejected. We explain this observation in remark \ref{rem:extrapolation}. \autoref{fig:sx5e_generative_performance_long_term_level} and \autoref{fig:sx5e_generative_performance_long_term_return} show the fitting performance of the level and return process respectively. Also here we observe that the fitting performance is accurate. Notably, we observe that the unconditional generated level distribution is slightly off from the historical, but still gives a good fit. Last, we also display the joint bi-variate distributions of the long-term return process in \autoref{fig:sx5e_kde_return} which shows that the calibrated flow gives an accurate fit. 

\begin{remark}[The extrapolation problem of network-based time series simulators]
\label{rem:extrapolation}
During simulation of the long time series we observed that with a non-zero probability some of the paths diverged / exploded. This observation is plausible: the simulator may generate states that it was not trained on or states that are off the empirical manifold. Since neural networks are known not to genalise well to unknown data (here past conditions $\bfy_t \in \Y$) (see for example \cite[Section 4.1]{arvanitidis2017latent}) it is natural, but highly unfavorable, to observe this phenomenon, as rejecting generated paths alters the distribution of the generated spot process causing it not to satisfy the martingale property which we wanted to guarantee by construction (see \autoref{sec:mono_martingale}). We believe that the extrapolation problem is challenging but essential for constructing high-fidelity market simulators and want to address it in future work. Last, we would like to note that this observation was not only made for our flow-based simulator but for any network-based simulator that was trained  using adversarial training techniques \cite{Hao2020, Wiese2019}. 
\end{remark}

\section{Multi-asset market simulation: sampling from joint distribution of the asset universe}
\label{sec:multi_simulator}
After introducing the conditional spot and equity option market simulator for a single underlying, we cover in this section the topic of simulating multi-asset markets. Assume an $N$-variate spot and equity option market and identify each underlying's individual market state with a superscript, i.e. $\bfx_t^i = (r_t^i, \boldsymbol{\sigma}_t^i), i = 1, \dots, N$, where $r_t^i$ and $\boldsymbol{\sigma}_t^i$ denote the $i$-th underyling's spot log-return and compressed DLV grid respectively. Let $T_\theta^i: \Y \to (\Z \stackrel{\cong}{\to} \X), i=1, \dots, N$ be \emph{optimal} simulators in the sense that for any asset $i=1, \dots, N$ the model densities $\px_\theta(\bfx^i_{t+1}|\bfy^i_t)$ are equal to the (unknown) target densities $ \px^*(\bfx^i_{t+1}|\bfy^i_t)$ for any $(\bfx^i_{t+1},\bfy^i_t) \in X \times Y $. Furthermore, denote by ${\bar{T}_\theta: \Y^N \to (\Z^N \stackrel{\cong}{\to} \X^N)}$ the \emph{joint market simulator} which is defined for any joint $p$-lagged market state $\bary_t = (\bfy_t^1, \dots, \bfy_t^N) \in \Y^N$ as 
\begin{equation*}
	\bar{T}_\theta(\barz_{t+1}; \bary_t)=\left(T^i_\theta(\bfz^i_{t+1}; \bfy^i_t)\right)_{i=1}^N
\end{equation*} 
where $\barz_{t+1} = (\bfz_{t+1}^1, \dots, \bfz_{t+1}^N) \in \Z$ is the joint latent process where $\bfz_{t+1}^i \sim \pz, i = 1, \dots, N$ follows the \emph{simple} source density which we assume to be the standard multivariate normal distribution $\mathcal{N}(0, I)$. 

Our aim in this section is to calibrate the joint distribution of the joint latent process in a scalable manner while not altering the conditional density of the individual simulators $T^i_\theta, i = 1, \dots, N$. We therefore assume for any joint $p$-lagged market state $\bary_t \in \Y^N$ that the joint density of the latent random variable is independent from the state
\begin{equation*}
	\label{eq:static_assumption}
	\tag{$\star$}
	\px^*(\barz_{t+1} | \bary_t) = \px^*(\barz_{t+1})
\end{equation*}
and refer to this as the \emph{static noise assumption}. Although the noise distribution may be state-dependent, our aim in this paper is to find a fast and scalable calibration algorithm for correlating the market simulators and we leave it as future work to construct state-dependent joint densities $\px^*(\barz_{t+1} | \bary_t)$ in a scalable manner.

\begin{figure}[h!]
	\centering
	\includegraphics[width=\textwidth]{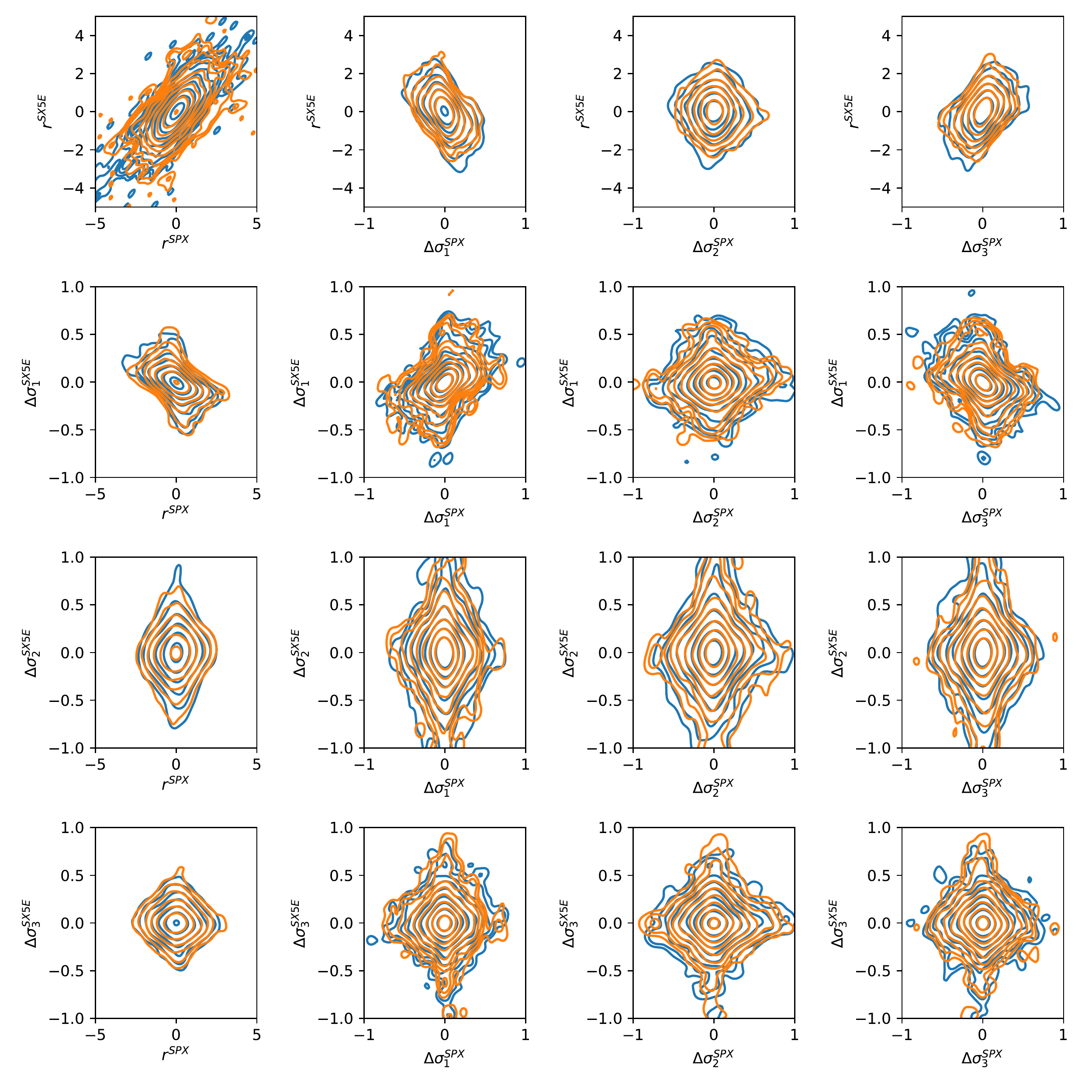}
	\caption{Eurostoxx 50 / S\&P 500 kernel density estimate of the historical (blue) and generated (orange) joint return distribution $(r_t^{\textrm{SX5E}}, \Delta \boldsymbol{\sigma}_t^{\textrm{SX5E}}, r_t^{\textrm{SPX}}, \Delta \boldsymbol{\sigma}_t^{\textrm{SPX}})$ where the generated joint latent distribution was governed by a Gaussian copula. We use the ticker symbols SX5E and SPX to abbreviate the Eurostoxx 50 and S\&P 500 components in the plot.}
	\label{fig:sx5e_spx_kde_return_gauss_copula}
\end{figure}

Under the static noise assumption it is sufficient to calibrate a copula \cite{nelsen2007introduction} to capture the multi-asset's joint distribution. For this purpose we propose to use a Gaussian copula to model the joint distribution. When specifying the Gaussian copula as the parametric family we assume that the joint latent process follows a normal distribution $\barz_{t+1} \sim \mathcal{N}(\mathbf{0}, \Sigma)$ with covariance matrix $\Sigma$ where the covariance matrix satisfies the block structure 
\begin{equation}
	\label{eq:sigma_block}
	\Sigma = \begin{pmatrix}
	  I_{d\times d} & \Sigma_{1, 2} &\cdots & \Sigma_{1, N} \\
	  \Sigma_{1, 2}^T & \ddots & \ddots & \vdots\\
	  \vdots & \ddots & \ddots & \Sigma_{N-1, N}\\
	  \Sigma_{1, N}^T & \cdots & \Sigma_{N-1, N}^T & I_{d\times d} 
	\end{pmatrix} \in \mathbb{R}^{Nd \times Nd}
\end{equation}
The covariance matrix can be estimated directly from the joint latent series
\begin{equation}
	\label{eq:sigma_estimate}
	\hat{\Sigma} = \dfrac{1}{T-p-1}\sum_{t=p+1}^{T}\left(\barz_t - \hat{\mu}_{\barz} \right) \otimes \left(\barz_t - \hat{\mu}_{\barz}\right)
\end{equation}
where $\hat{\mu}_{\barz}$ denotes the sample mean and $\otimes$ is used to denote the outer product. To ensure that the dynamics of each individual simulator are not changed through sampling of the Gaussian copula, we additionally enforce during estimation of $\Sigma$ that the blocks along the diagonal are equal to the identity matrix. Thus, if $\barz_{t+1} \sim \mathcal{N}(0, \hat{\Sigma})$ and $\hat{\Sigma}$ is of the form \eqref{eq:sigma_block}, then the conditional model densities $q_\theta(\bfx_{t+1}^{i}| \bfy_{t}^{i}), i = 1, \dots, N$ remain unchanged.  

We benchmark the Gaussian copula against a more complicated flow-based model $S_\theta: \U \to \Z$, as described in \autoref{sec:normalizing_flows_background}, that will aim to transform a standard normal source distribution to the targeted distribution $\barz\sim\px^*$. It is important to note that while the flow is able to capture non-linear structure of the joint distribution of the joint latent time series, it also alters the marginal density resulting in a changed model density $q_\theta(\bfx_{t+1}^{i}| \bfy_{t}^{i}), i = 1, \dots, N$. While it is possible to account for this change in the marginal density, it is non-trivial to say how strong the effect will be on the calibrated joint density after such a correction. 

\begin{figure}[!h]
	\centering
	\includegraphics[width=\textwidth]{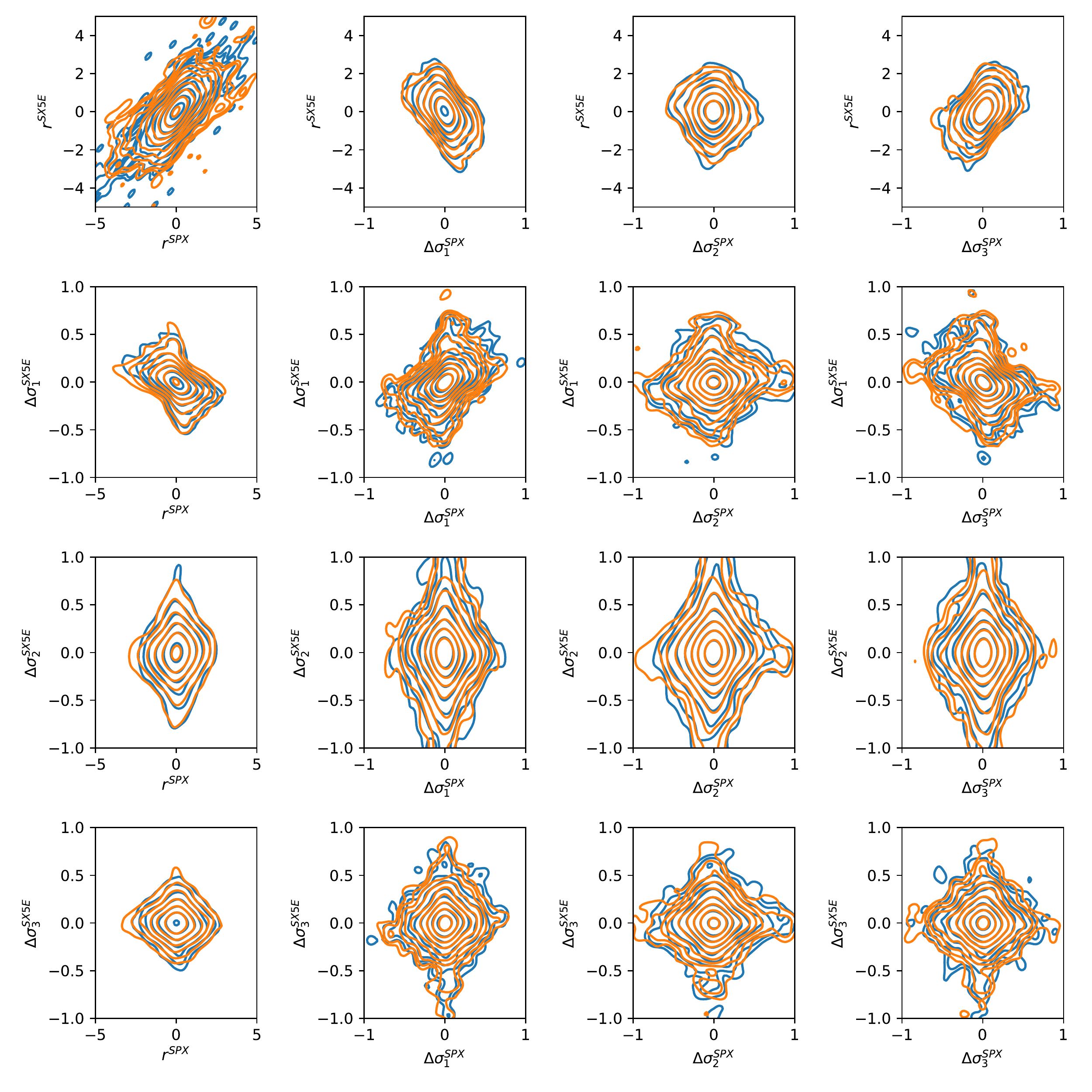}
	\caption{Eurostoxx 50 / S\&P 500 kernel density estimate of the historical (blue) and generated (orange) joint return distribution $(r_t^{\textrm{SX5E}}, \Delta \boldsymbol{\sigma}_t^{\textrm{SX5E}}, r_t^{\textrm{SPX}}, \Delta \boldsymbol{\sigma}_t^{\textrm{SPX}})$ where the generated joint latent distribution was governed by a normalizing flow.}
	\label{fig:sx5e_spx_kde_return_flow}
\end{figure}

\subsection{Numerical results: Eurostoxx 50 and S\&P 500}
Above we motivated the use of copulas for modeling the joint latent distribution. In this section, we benchmark the Gaussian copula against a flow-based calibration. We consider the Eurostoxx 50 as well as the S\&P 500 from January 2011 to November 2021 for the same DLV grid we considered in \autoref{sec:mono_numerical}, i.e. maturities $\lbrace 20, 40, 60, 120 \rbrace$ (quoted in business days) and relative strikes $\lbrace 0.8, 0.85, \dots, 1.2 \rbrace$. Prior to calibration of the joint latent distribution we proceed as in \autoref{sec:mono_compression} and \autoref{sec:mono_numerical} to calibrate the autoencoder with a three-dimensional bottleneck, realized volatility estimator $T_r$ and compressed state simulator map $T_{\boldsymbol{\sigma}}$ for each underlying individually. 

After calibration of the market simulators we compute the latent process by inverting the flow. We then estimate the covariance matrix $\Sigma \in \mathbb{R}^{2d \times 2d}, d=4$ through \eqref{eq:sigma_estimate} and impose the block structure as described in \eqref{eq:sigma_block}. Afterwards, we train the unconditional flow. Here, the flow will be represented by a three layer network with 64 hidden dimensions. Similar to the procedure described in \autoref{sec:mono_numerical} the data is split into a train and test set, the Adam optimizer is used with a learning rate of $0.001$ and $10\%$ DropOut is used. 

\begin{figure}[h!]
	\centering
	\includegraphics[width=\textwidth]{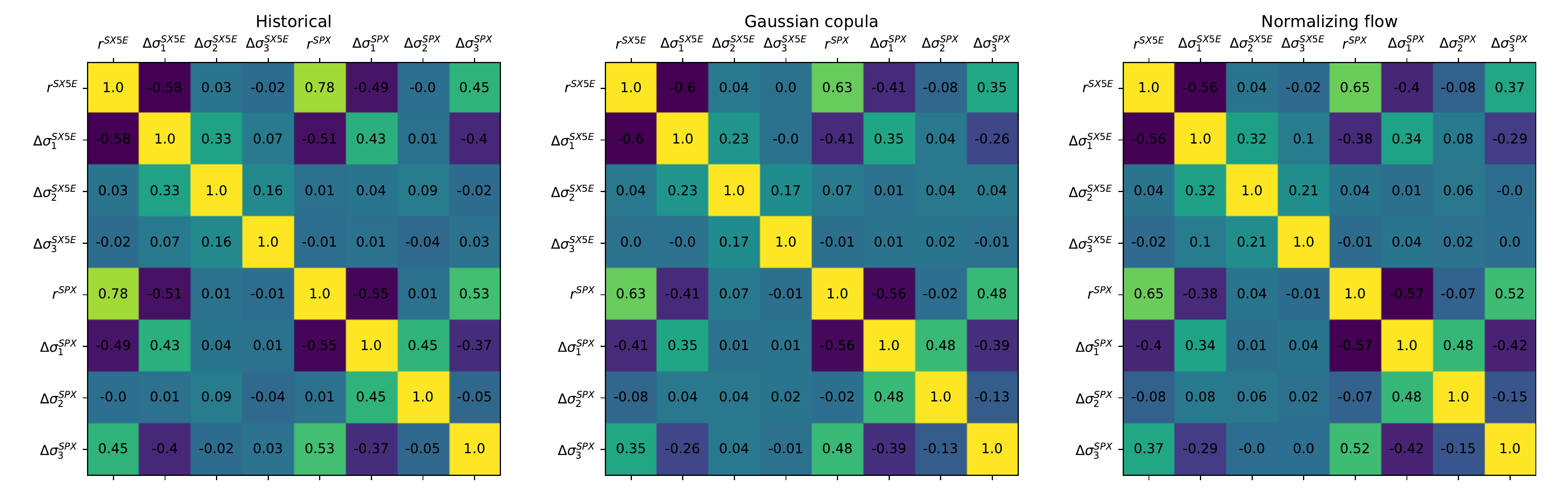}
	\caption{Eurostoxx 50 / S\&P 500 cross-correlation matrix of the joint return process of the historical data (left) and generated data under the Gaussian copula (middle) and normalizing flow (right).}
	\label{fig:sx5e_spx_return_cross_correl}
\end{figure}

Afterwards, evaluate the fitting performance and sample paths on a short time horizon by using the sampling scheme described in \autoref{sec:mono_numerical}. To assess the fitting performance we perform a kernel density estimate on the bivariate joint distributions. The kernel estimate of the historical (blue) and generated (orange) return distribution is depicted in \autoref{fig:sx5e_spx_kde_return_gauss_copula} for the Gaussian copula and in \autoref{fig:sx5e_spx_kde_return_flow} for the flow-based copula. Furthermore, \autoref{fig:sx5e_spx_return_cross_correl} shows the cross-correlation matrix of the return distribution of the historical, Gaussian copula simulated and flow simulated paths. Comparing both visually the kernel density estimate of the Gaussian copula and the normalizing flow it is not apparent which of the two outperforms the other. A similar claim can be made when looking at the cross-correlation matrices as both matrices do not deviate strongly from another. Since the Gaussian copula ensures that the dynamics of the individual market simulators remain unchanged, we conclude in this numerical example that the Gaussian copula is more favorable than the flow-based calibration.    

\section{Conclusion and future research}
\label{sec:conclusion}
In this paper, we presented formally a novel spot and equity option market simulator for a single underlying by utilizing normalizing flows. To overcome the curse of dimensionality of high-dimensional call price lattices we introduced an autoencoder which was able to learn efficient low-dimensional representations while maintaining no static arbitrage in the reconstructed surface by construction. We then introduced a fast and scalable method for calibrating the joint distribution of multiple independent simulators that were calibrated on distinct underlyings. We end this paper with future avenues of research / work. 

\paragraph{Extrapolation problem}
In \autoref{sec:mono_numerical}, we observed that with non-zero probability paths had to be rejected due to path explosions. Unfavorably, this has an effect on the generated spot distribution causing it to not satisfy the martingal property. While it is difficult to discover a regularization method that counteracts such explosions we find it to be a central research topic and want to advance this area with future work. 

\paragraph{Realized volatility}
We assumed in \autoref{sec:mono_martingale} that realized volatility is a function of past market states. This estimator can be improved by simply replacing it through an estimator that estimates realized volatility $\nu_t$ from intraday prices / tick data. Furthermore, the market simulator can be improved by augmenting the market state through the realized volatilty estimate $\bfx_t = (r_t, \boldsymbol{\sigma}_t, \nu_t)$ such that the dynamics of $\nu_t$ are modeled within the market simulator. 

\paragraph{High-fidelity conditional spot distributions}
To ensure the martingale property in the generated spot distribution a normal form \eqref{eq:normal_form} was imposed on the log-returns distribution in \autoref{sec:mono_martingale}. In general, the constraint is restrictive as it cannot capture features such as the skew and tail of the historical distribution. We believe it is fruitful to incorporate jump processes for modeling the generated spot distribution to more accurately capture tails and generalise the normal form, perhaps through a normalizing flow-based construction which satisfies the martingale property.

\clearpage
\printbibliography

\clearpage
\appendix
\section{Disclosure}
Opinions and estimates constitute our judgement as of the date of this Material, are for informational purposes only and are subject to change without notice. It is not a research report and is not intended as such. Past performance is not indicative of future results. This Material is not the product of J.P. Morgan’s Research Department and therefore, has not been prepared in accordance with legal requirements to promote the independence of research, including but not limited to, the prohibition on the dealing ahead of the dissemination of investment research. This Material is not intended as research, a recommendation, advice, offer or solicitation for the purchase or sale of any financial product or service, or to be used in any way for evaluating the merits of participating in any transaction. Please consult your own advisors regarding legal, tax, accounting or any other aspects including suitability implications for your particular circumstances. J.P. Morgan disclaims any responsibility or liability whatsoever for the quality, accuracy or completeness of the information herein, and for any reliance on, or use of this material in any way. Important disclosures at: \href{www.jpmorgan.com/disclosures}{\texttt{www.jpmorgan.com/disclosures}}.

\section{Further numerical results}
\label{sec:numerical_results_appendix}
\begin{figure}[h!]
	\centering
	\includegraphics[width=\textwidth]{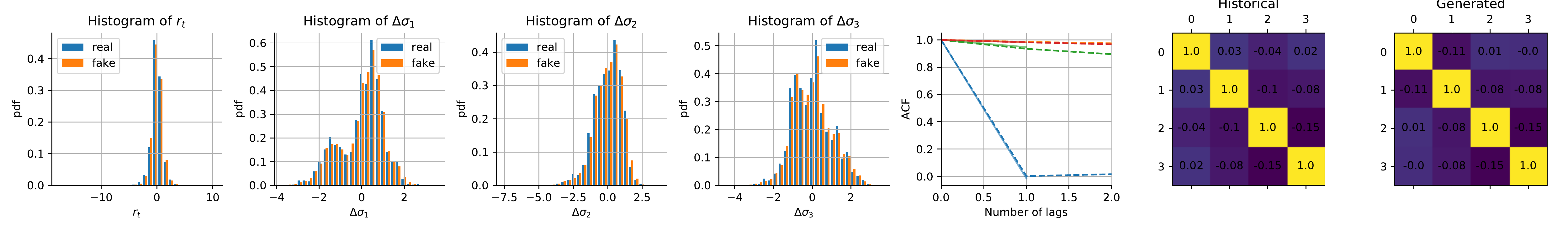}
	\caption{Eurostoxx 50 short-term performance of the level process $(r_t, \boldsymbol{\sigma}_t)$. From left to right: histograms of $r, \boldsymbol{\sigma}_{t, 1}, \boldsymbol{\sigma}_{t, 2}, \boldsymbol{\sigma}_{t, 3}$ where blue indicates the color of the historical dataset and orange the color of the generated, ACF of the level process for the first two lags, and last the cross-correlation matrix of the historical (left) and generated (right).}
	\label{fig:sx5e_generative_performance_short_term_level} 
\end{figure}

\begin{figure}[h!]
	\centering
	\includegraphics[width=\textwidth]{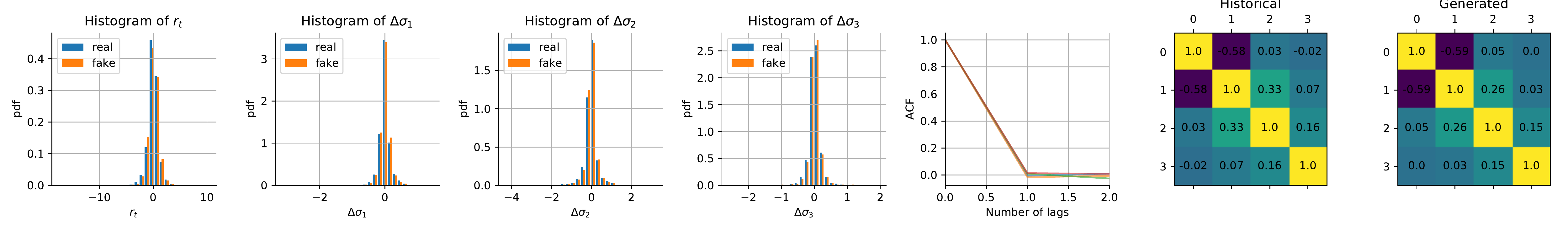}
	\caption{Eurostoxx 50 short-term performance of the return process $(r_t, \Delta \boldsymbol{\sigma}_t)$. From left to right: histograms of $r_t, \Delta \boldsymbol{\sigma}_{t, 1}, \Delta \boldsymbol{\sigma}_{t, 2}, \Delta \boldsymbol{\sigma}_{t, 3}$ where blue indicates the color of the historical dataset and orange the color of the generated, ACF of the return process for the first two lags, and last the cross-correlation matrix of the historical (left) and generated (right).}
	\label{fig:sx5e_generative_performance_short_term_return} 
\end{figure}

\end{document}